\documentclass[aps,showpacs,nofootinbib]{revtex4-1}
\usepackage{amsmath}
\usepackage{amsfonts}
\usepackage{amssymb,amscd,amsbsy}
\usepackage{epsfig}
\usepackage{color}
\usepackage{graphicx}
\usepackage{subcaption}

\usepackage{mathtools}

\newcommand{\bea}{\begin{eqnarray}}
\newcommand{\eea}{\end{eqnarray}}

\newcommand{\vect}[1]{\mathbf{#1}}

\newcommand{\tr}{\textcolor{black}}
\newcommand{\kt}{k_{\rm B}T}

\newcommand{\cref}{c^{(2)}_{\rm ref}}

\newcommand{\rhon}{\rho_{\rm nem}}
\newcommand{\rhoi}{\rho_{\rm iso}}
\newcommand{\zc}{z_{\rm coex}}
\newcommand{\etac}{\eta_{\rm coex}}
\newcommand{\Qmax}{Q_{\rm max}}
\newcommand{\Qmin}{Q_{\rm min}}

\begin{document}
\title{The isotropic--nematic transition for hard rods on a three--dimensional (3D) cubic lattice}

\author{A. Gschwind$^{1}$, M. Klopotek$^{1}$, Y. Ai$^{2}$, and M. Oettel$^{1}$}
\affiliation{
$^1$ Institut f\"ur Angewandte Physik, Eberhard Karls Universit\"at T\"ubingen, D--72076 T\"ubingen, Germany \\
$^2$ School of Engineering and Applied Science, University of Pennsylvania, Philadelphia, PA 19104, USA 
}

\email{martin.oettel@uni-tuebingen.de}

\begin{abstract}
Using grand canonical Monte Carlo (GCMC) simulations, we investigate the  isotropic--nematic phase transition for hard rods
of size $L \times 1 \times 1$ on a 3D cubic lattice. We observe such a transition for $L \ge 6$. For $L=6$, the nematic state
has a negative order parameter, reflecting the co--occurrence of two dominating orientations. For $L \ge 7$, the nematic state
has a positive order parameter, corresponding to the dominance of one orientation. We investigate rod lengths up to 
$L=25$ and find evidence \tr{for a very weakly first-order isotropic-nematic transition, while we cannot completely
 rule out a second order transition.}
It was not possible 
to detect a density jump at the transition, despite using large systems containing several $10^5$ particles. 
The probability density distributions $P(Q)$ from the GCMC simulations \tr{near the transition} are very broad, 
pointing to
strong fluctuations. Our results complement earlier results on the demixing (pseudonematic) transition for an equivalent system in 2D,
which is presumably of Ising--type and occurs for $L \ge 7$. We compare our results to lattice fundamental measure theory (FMT) 
and find that FMT strongly overestimates nematic order and consequently predicts a strong first order transition.
The rod packing fraction of the nematic coexisting states, however, agree reasonably well between FMT and GCMC.  
\end{abstract}

\maketitle

\section{Introduction}

Investigating phase transitions on lattice models has often helped 
formulate general concepts in phase transition theory, if not initiate them. Additionally, 
lattice models often take up the role of a simplified version of a continuum model of interest, in the hope
they retain the basic physics, e.g. the type of phase transition. Popular examples are using the lattice gas 
to explore the liquid--vapor transition or lattice polymer models in order to reduce the large
configurational space accompanying their continuum progenitors, enabling specific questions to be addressed.

Systems of hard, anisotropic particles have found steady interest over the past decades
since they exhibit many types of phase transitions seen in molecular or colloidal materials, while being purely entropic in nature:
These include transitions from
an isotropic liquid phase to various smectic and crystalline phases via 
nematic and biaxial states \cite{Mar14}. 
Not only can hard--body models in the continuum  be  treated efficiently
in the context of simulations~\cite{All93}, but also advanced density functional methods
can describe them accurately (e.g. fundamental measure theory) \cite{Mar14,Witt_thesis15}.
It is hence interesting to see whether simple lattice analogs of these models show equivalent phase transitions, and likewise are  
accessible to density functional methods. In two dimensions (2D), hard rods with side--lengths $L \times 1$ 
on a square lattice define the most basic lattice model; in 3D, the corresponding lattice model entails
hard rods with side--lengths $L \times 1 \times 1$ on a cubic lattice.

In 2D, several simulation studies have investigated the basic model introduced above.
One can view it as a binary system of particles (rods oriented in $x$-- and $y$--direction, respectively), thus
a transition to a
bulk state with preferred orientation corresponds to a demixing transition presumably of Ising type.    
This demixing is found for $L \ge 7$  in simulations~\cite{Ghosh07}. The critical packing fraction for the onset of demixing scales approximately
as \tr{$4.8/L$ for large $L$~\cite{Mat08,Kun16}}.
At very high packing fractions $\eta \approx 1$, theoretical arguments predict a reentrant
transition from the demixed to a disordered state bearing some characteristics of a cubatic phase
on a lattice \cite{Ghosh07}. This transition has been studied in more detail using simulations in Refs.~\cite{Kun13,Kun14}.
For rods with extensions $mL \times m$
(where $mL,m$ are integer and $L$ may be noninteger), the phase diagram has been investigated in Refs.~\cite{Kun15,Kun16}
where (for $m>1$) it is shown that a columnar phase appears between the demixed and high-density disordered phase.

Theoretical studies of the bulk phase of the basic models in 2D and 3D were sparked by DiMarzio~\cite{DiMar61}, having approached the problem from the 
context of polymer theory. 
DiMarzio calculated the number of possible packings of rods---thus evaluated the entropy---through
approximating the probability of inserting a new rod into a system already containing other rods. 
The approximation consists of treating all rods orthogonal
to the rod that is to be inserted in a typical mean--field fashion: one such orthogonal rods (of length $L$) is considered a `cloud' of $L$ 
independently--distributed obstacles (of size $1 \times 1$ in 2D and $1 \times 1 \times 1$ in 3D) for the rod to be inserted. DiMarzio,
however, did not evaluate the bulk state explicitly by maximizing the entropy. This appears to have first been done 
by Alben \cite{Alb71} for the 3D model, who found a strong first order nematic transition for rod lengths $L \ge 4$.    
The equivalence of the DiMarzio entropy to exact solutions on Bethe--like lattices has been investigated in Ref.~\cite{Dhar11} 
(see also the much earlier Ref.~\cite{Cot69} regarding such a connection). Among other results, Alben's solution is recovered, and
the 2D solution is given: demixing of the two rod species occurs for
$L \ge 4$ above a  critical packing fraction $\eta_{\rm c}(L) = 2/(L-1)$. 
This is in contrast to the aforementioned simulations \cite{Ghosh07}, which found demixing for $L \ge 7$.

A completely different theoretical route for treating hard particles on lattices has been initiated by 
Lafuente and Cuesta~\cite{Laf02,Laf04} using Fundamental Measure Theory (FMT). They provide a general recipe for
how to construct a density functional for arbitrary hard particle mixtures, applicable to inhomogeneous situations.
The construction of the functional begins with the requirement that the functional is exact for extreme confinement in so--called zero--dimensional
cavities. A zero--dimensional 
cavity is a restricted domain on the lattice which
can only hold one particle at a time and whose free energy is known. Such a cavity may
consist of more than one point where the hard particle is positioned.
For the basic rod models in 2D and 3D, the FMT functional and its bulk properties have been worked out in Ref.~\cite{Oet16}. 
Surprisingly, it turns out that the bulk entropy is exactly the same as DiMarzio's (which was not mentioned in \cite{Oet16}), as well as
that for rods on Bethe-like lattices with coordination numbers 4 and 6 (representing 2D and 3D, respectively)~\cite{Dhar11}; 
therefore, FMT likewise predicts a strong first
order transition from an isotropic to a nematic state. The
relative coexistence gap $2(\rhon-\rhoi)/(\rhon+\rhoi)$ ($\rhon$ and $\rhoi$ are the coexistence number densities
per lattice site in the nematic and isotropic state, respectively) is narrow for shorter $L$ (e.g., 2.5\% for $L=4$), but 
quickly widens, becoming about 40\% in the limit of very long rods. This is in line with the general expectations 
from Landau theory, which predicts (by means of general symmetry arguments) a first order liquid--nematic transition in 3D for
hard bodies \cite{vanRoij16}.   

It is  interesting, therefore, to see whether simulations of the basic rod model in 3D can confirm this type of transition.
This is the purpose of the present work. To our surprise, we have not found studies in the literature addressing the
nematic transition for this basic 3D rod model. In Ref.~\cite{Case95}, rods with size $15 \times 3 \times 3$ were studied
and a layering transition was found.    
As shown below, our grand canonical simulations for rod lengths $L$ between 5 and 25 indicate that the isotropic--nematic phase
transition sets in for $L=6$ (resembling the layering transition of Ref.~\cite{Case95}), 
 moreover changes its character for $L \ge 7$ (explained below), and, rather peculiarly, is possibly continuous or only of very weak first--order type up to $L=25$, 
contrary to the theory above.  
The remainder of the paper is structured as follows: In Sec.~\ref{sec:model}, we introduce the model and its order parameters.
In Sec.~\ref{sec:methods}, the application of grand canonical Monte Carlo simulations for the model is described, and the FMT bulk 
free energy is recapitulated. The simulation results are described in Sec.~\ref{sec:results} and compared to FMT. A summary
and a discussion of how the  present results relate to those for continuum models is given in Sec.~\ref{sec:summary}.  
In the Appendix, the qualitative analogy of the 3D hard rod model with the three--state Potts model is exploited, and
an investigation with  an appropriate order parameter is done for $L=8$, including finite--size analysis.   

\section{Model}
\label{sec:model}

We consider a simple cubic lattice in 3D, the unit cell length is set to 1. Hard rods are parallelepipeds with 
extensions $L \times 1 \times 1$ and corners sitting on lattice points. The position of a rod is specified by the corner 
whose lattice coordinates are minimal each. They are allowed to touch (i.e., share corners or faces),
but forbidden to overlap. The cubic lattice restricts possible orientations to three, and we refer to rods oriented
in $x$-- resp. $y$-- resp. $z$--direction as species 1 resp. 2 resp. 3. Species densities $\rho_i$ are defined as number of rods of
species $i$ ($N_i$) per lattice site, $\rho=\rho_1 + \rho_2+\rho_3$ is the total density and $\eta=\rho L \le 1$ is the total packing
fraction. We introduce bulk nematic order parameters $Q_i$ and biaxiality order parameters
$S_i$ by
\bea
 \label{eq:Qidef}
   Q_i & = &  \frac{\rho_i - \frac{\rho_j +\rho_k}{2}}{\rho} \;, \\
 \label{eq:Sidef}
   S_i &=  &  \frac{\rho_j -\rho_k}{\rho_j+\rho_k} \;,
\eea 
where $(ijk)$ is a cyclic permutation of $(123)$.

An allowed configuration of the system is given by the set of all rod positions ${\mathcal S}=\{\vect s_{j,k}\}$ where 
$j=1,2,3$ indicates the species and $k=1 ... N_j$ the rod number and in which the rods do not overlap.
The total number of allowed configuations is denoted by ${\mathcal N}(\{N_j\})$.
The grand partition function of the system is defined by
\bea
  \Xi (z_i) = \sum_{N_1=0}^\infty \sum_{N_2=0}^\infty \sum_{N_3=0}^\infty \frac{z_1^{N_1} z_2^{N_2} z_3^{N_3}}{N_1! N_2! N_3!}
  \; {\mathcal N}(\{N_j\}) \;,
\eea
where $z_i =\exp(\beta\mu_i)$ is the activity of species $i$ ($\beta=1/(\kt)$ is the inverse temperature and
$\mu_i$ is the chemical potential of species $i$). In a bulk system all $z_i$ are equal, $z_i=z$. 
Note that the 3 possible particle orientations can be treated as an internal property of particles  such that
the grand partition function in the bulk can  be written alternatively as a single sum over the total number of particles:
\bea
  \Xi(z) = \sum_{N=0}^\infty \frac{z^N}{N!} \;  {\mathcal N'}(N) \;,
\eea
where ${\mathcal N'}(N)$ is the total number of allowed configurations having $N$ particles with three
possible orientations each. The grand canonical average of an observable $A(\{N_j\})$ is defined by
\bea
  \langle A \rangle  = \frac{1}{\Xi} \sum_{N=0}^\infty \frac{z^N}{N!} \sum_{\mathcal S} A(\{N_j\})  \;.
\eea

\section{Methods}
\label{sec:methods}

\subsection{Grand canonical Monte Carlo (GCMC) simulations}

We have simulated the model on cubic lattices with size $V=M^3$. 
We have worked with lattice sizes between {$M=50$ and $M=170$}, depending on rod length. 
In each step, a  particle insertion attempt or deletion attempt
was chosen with probability $1/2$. For an insertion move, the orientation and the position of the particle
was picked randomly. For the deletion move, one of the particles in the system was chosen randomly. Detailed balance was obeyed: the insertion move was accepted with a probability ${\rm min}(1, 3z V/(N+1))$ if it lead to
an allowed configuration, and the deletion move was accepted with a probability \tr{${\rm min}(1, N/(3zV) )$}. 
$N$ indicates the number of particles before the insertion/deletion move was attempted.

To compute densities and orientation variables, we stored various ``time'' series of the rod numbers $N_i(t)$,
measured every $10^7$ moves (otherwise noted). In isotropic states, $N_i(t)$ fluctuated around $\langle N \rangle /3$, and
equilibrium was  reached promptly. On the other hand, one species having a pronounced majority was a typical sign of being in e.g. the nematic state; there,
intervals of alternating majority species were observed in the time series. We assumed full equilibration 
of the system if of the order of 10 such intervals occurred. This criterion was of course
difficult to fulfill deep in the nematic phase. The collected time series were used to calculate averages $\langle A \rangle$ and probability density distributions
$P(A)$ (histograms). 

The signature of a first--order transition in a GCMC simulation is usually a double peak in the probability density distribution 
$P(N)$, and the peaks should have equal area at coexistence (activity $\zc$).
In the vicinity of coexistence between an isotropic and a nematic state with excess of {\em one} species,
the probability density distribution of one of the nematic order parameters (say $P(Q_1)$) should furthermore exhibit three peaks: 
one peak centered at $Q_1 \approx 0$ and
two peaks centered at $Q_1=q>0$ and $Q_1=-q/2$, where the second peak corresponds to the
species 2 or 3 being the majority species. Consider the distribution $P(\Qmax)$
with $\Qmax = \max\limits_i (Q_i)$: it should exhibit a double--peak structure located at
$\Qmax \approx 0$ and $\Qmax = q>0$. On the other hand, a transition to a nematic state with excess of {\em two} species
is best observed via the distribution $P(\Qmin)$ with $\Qmin = \min\limits_i (Q_i)$.
It should likewise exhibit a double--peak structure located at
$\Qmin \approx 0$ and $\Qmin = q<0$. 

A negative nematic order parameter in continuum models for rods would correspond to rods preferentially orienting
perpendicular to a director. This has not been reported in the literature in the case of uniaxial hard rods.

\subsection{Lattice FMT}

A general FMT functional for hard rod mixtures on lattices with arbitrary dimensions has been derived by 
Lafuente and Cuesta~\cite{Laf02,Laf04}. For the specific case of rods with length $L$ and width 1,  
the basic definitions and examples 
for the functionals and their equilibrium properties in two and three dimensions
are provided in Ref.~\cite{Oet16}.
We only need the free energy density in the homogeneous case (bulk) for the present work. 
\bea
    f &=& f^{\rm id} + f^{\rm ex}   \qquad \mbox{with}\label{eq_fdef} \\
    \beta f^{\rm id} & =& \sum_{i=1}^3 \rho_i \ln\rho_i - \rho \;,\label{eq_fiddef} \\
     \beta f^{\rm ex} &=& \Phi^{\rm 0D}(L\rho) -  \sum_{i=1}^3 \Phi^{\rm 0D} \left( (L-1)\rho_i\right) \;.
\eea
Here, $\beta =1 /(\kt)$ is the inverse temperature which is set to 1 and
\bea
    \Phi^{\rm 0D}(\eta) &=& \eta + (1-\eta) \ln(1-\eta) \;.
\eea
is the excess free energy of a zero--dimensional cavity (which can hold no or only one particle) depending 
on its average occupation $\eta\in[0,1]$. The equilibrium bulk state is found by minimizing $f$ at constant total density $\rho$ with respect to the order parameters
$Q \equiv Q_i$ and $S \equiv S_i$ (for one specific $i$). We have found stable states only for $S=0$ and $Q \ge 0$. Isotropic--nematic 
coexistence is determined by equating the chemical potential $\mu=\mu_1=\mu_2=\mu_3$ (with $\mu_i=\partial f/\partial \rho_i$)
and pressure $p=\mu\rho-f$ between the isotropic and nematic states. 

\clearpage
\section{Results}
\label{sec:results}

 
\subsection{GCMC results}

For all rod lengths up to $L=25$, we did not see any sign of a 
double peak in $P(N)$, indicating that a hypothetical density difference between isotropic and nematic states would be  well below 1\%.

For $L=5$, the distributions $P(\Qmin)$ and $P(\Qmax)$ show a single peak near zero for packing fractions up to
\tr{0.84 ($z=25$)}, indicating a stable isotropic phase. \tr{For higher packing fractions, our simple algorithm did not equilibrate the system
on a reasonable time scale. Therefore, we cannot exclude the possibility for a nematic phase of some sort at higher packing fraction.}

\begin{figure}[h]
\includegraphics[width=7cm]{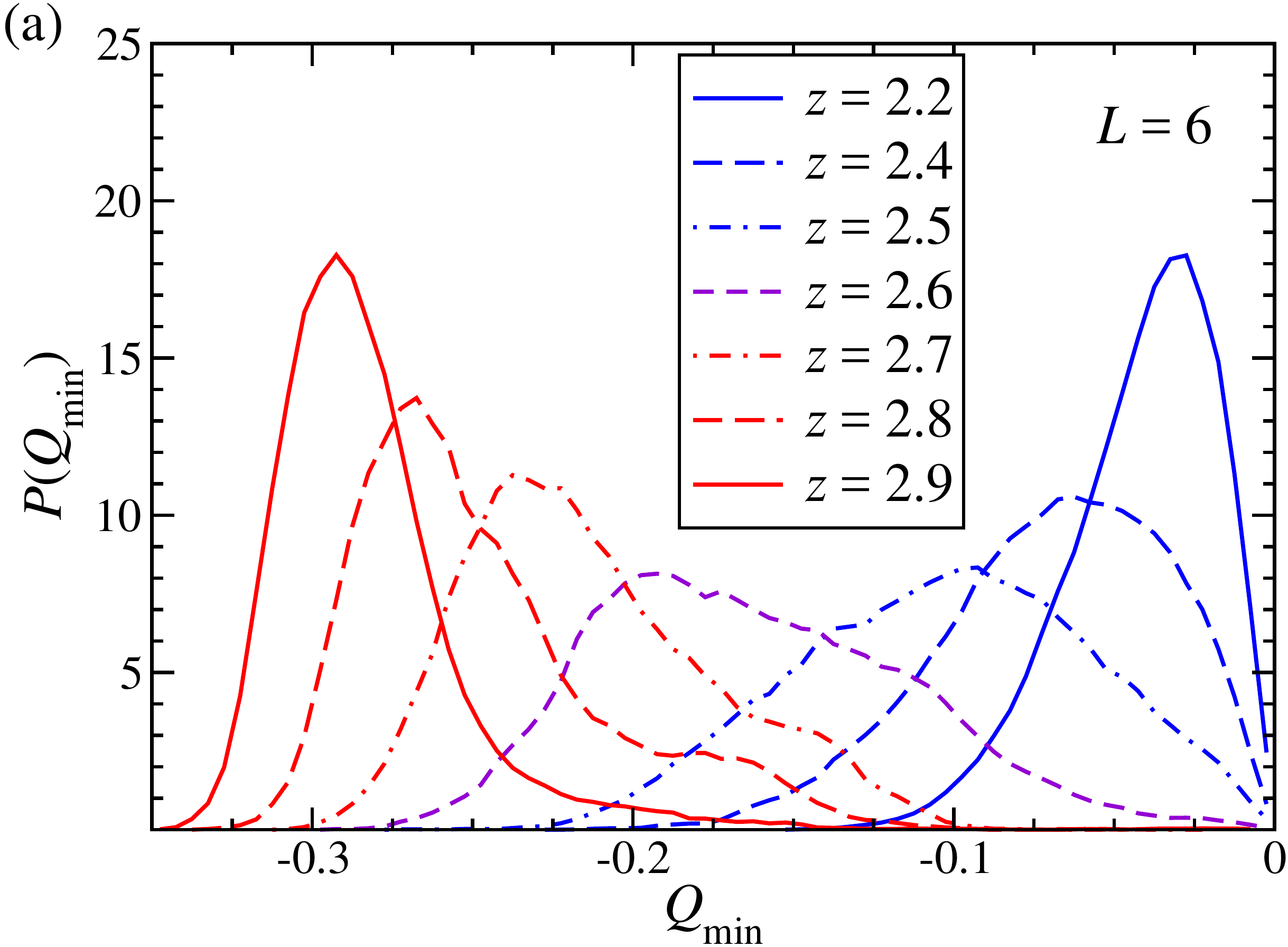}\hspace{1mm}
\includegraphics[width=7cm]{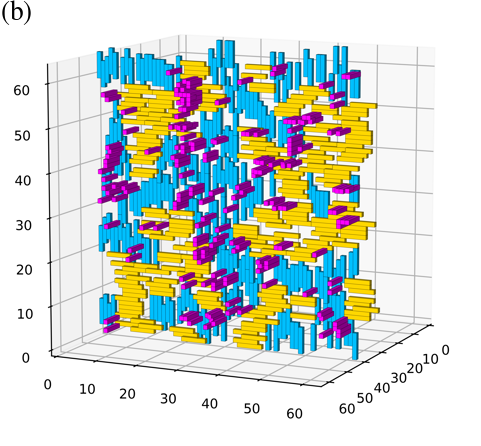}\\[3mm]
\includegraphics[width=\linewidth]{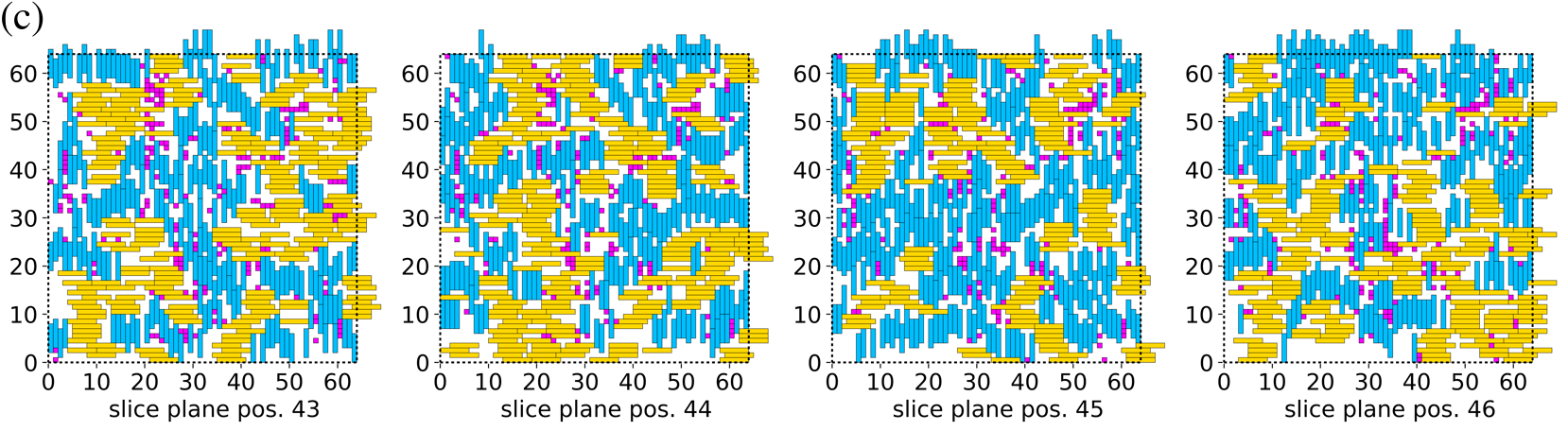}
 \caption{(a) The distribution $P(\Qmin)$ from GCMC for $L=6$ ($M=75$, bin size $\delta Q=0.005$) 
and for several activity values between  $z=2.2$ (isotropic state),
$z=2.6$ (near coexistence) and $z=2.9$ (nematic state). Near coexistence the maximal relative error estimates in the middle of the histogram are (posssibly exaggerated) at $\sim40\%$, attributed to by $\Qmin$ decorrelating very slowly. 
(b) Illustration: A slice through the preferred plane in the nematic phase with negative order parameter.
(c) Slices in the preferred plane at subsequent lattice points on the plane normal:
   one sees very few particles of the minority species and the particle positions of the majority species
  (patches of horizonal and vertical particles) do not correlate between the subsequent layers,
  thus the slices approximately correspond to independent 2D rod systems. 
  Snapshots are for {$z=3.8$, $M=64$}.}
 \label{fig:L6}
\end{figure}

For $L=6$, upon varying $z$ between 2.2 and 2.9, $P(\Qmin)$ transforms from one fairly sharp peak near zero, via a very broad peak
(resulting from an overlap of two broad peaks), and finally
to a moderately sharp peak at $\Qmin <0$. This indicates a transition from an isotropic state to a
nematic state with two excess species (negative order parameter). Results for $P(\Qmin)$ for several $z$
(illustrating this transition) are shown in 
Fig.~\ref{fig:L6}(a), and a series of snapshots in Fig.~\ref{fig:L6}(c), which display cuts in the plane defined by the two excess species
for $z=3.8$ (in the nematic phase). The snapshots show that the system separates into weakly coupled layers that are 
essentially only populated by the excess species (i.e. 2D systems). Within these layers, there is no dominance of one of the
excess species, in accordance with the absence of demixing for rods with $L=6$ in 2D \cite{Ghosh07}.  We note that $\Qmin$ (or $\Qmax$)---in contrast to $N$---is particulary sensitive to incomplete thermalization of the system (i.e. insufficient exploration of all possible equilibrium configurations), which happens with the standard algorithm used once this layering occurs.  

\begin{figure}[h]
	\begin{subfigure}[b]{7cm}
		\phantomsubcaption
		\includegraphics[width=\textwidth]{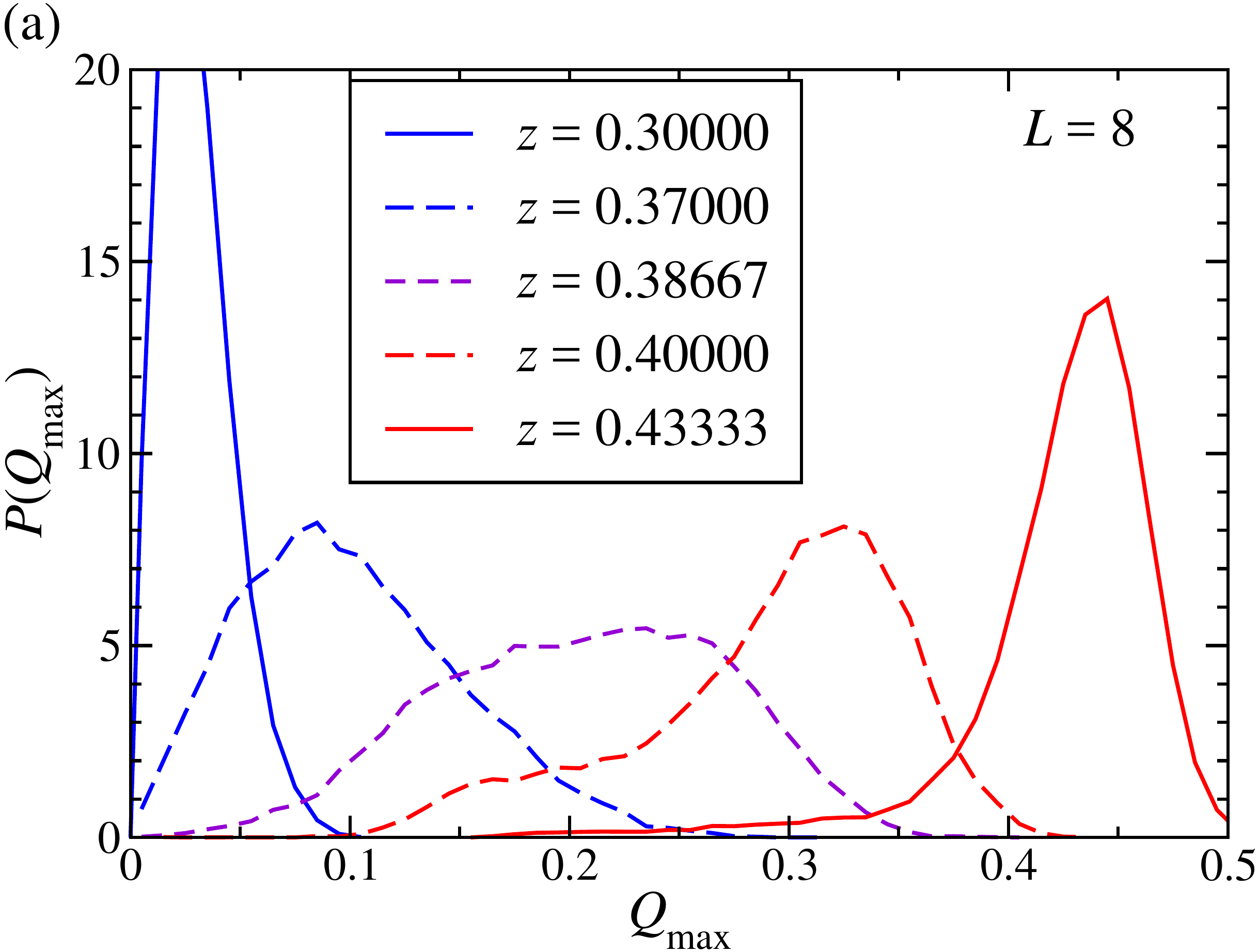}
		\label{fig:2a}
	\end{subfigure}
	\begin{subfigure}[b]{7cm}
		\phantomsubcaption
		\includegraphics[width=\textwidth]{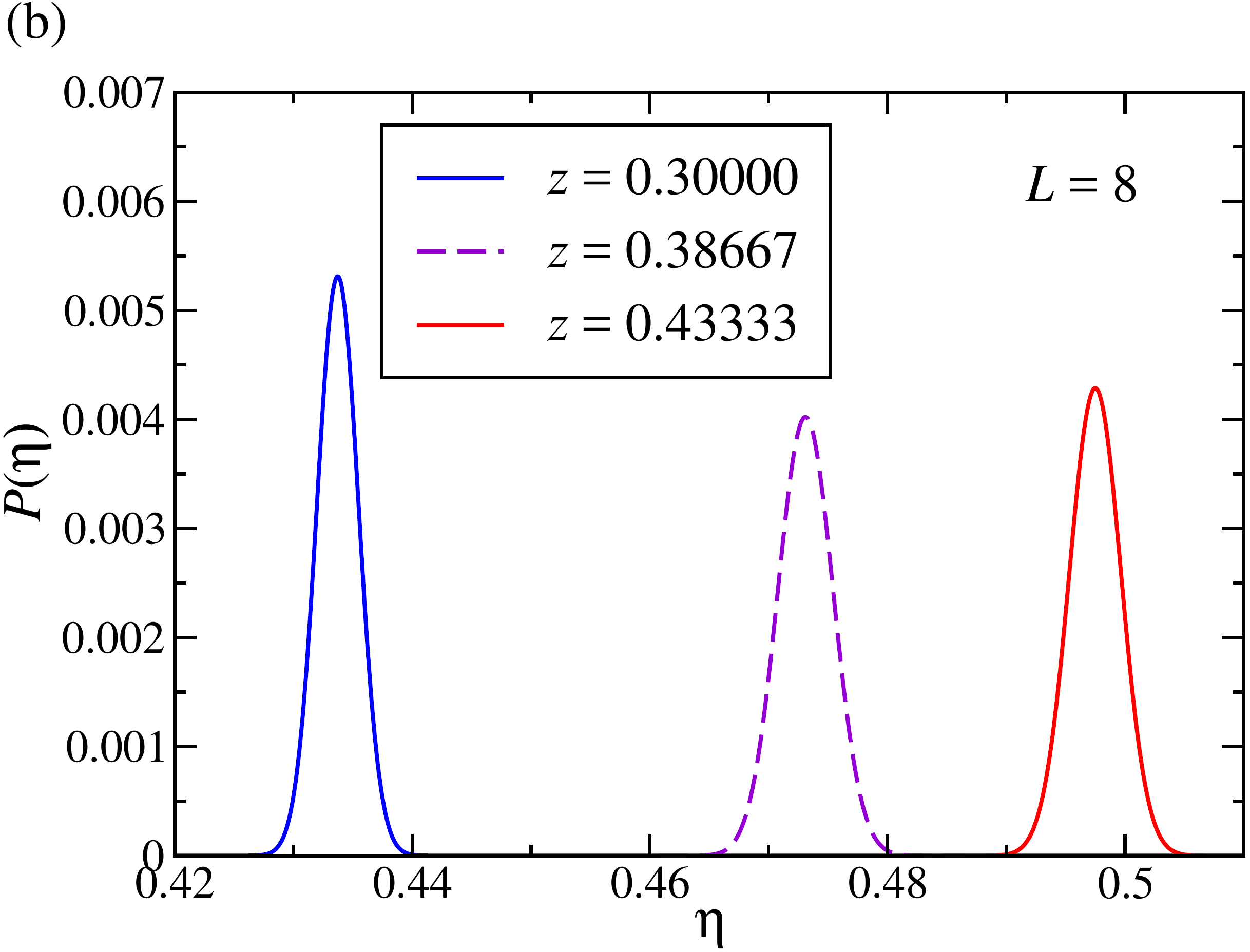}
		\label{fig:2b}
	\end{subfigure}
	\begin{subfigure}[b]{7cm}
		\phantomsubcaption
		\includegraphics[width=\textwidth]{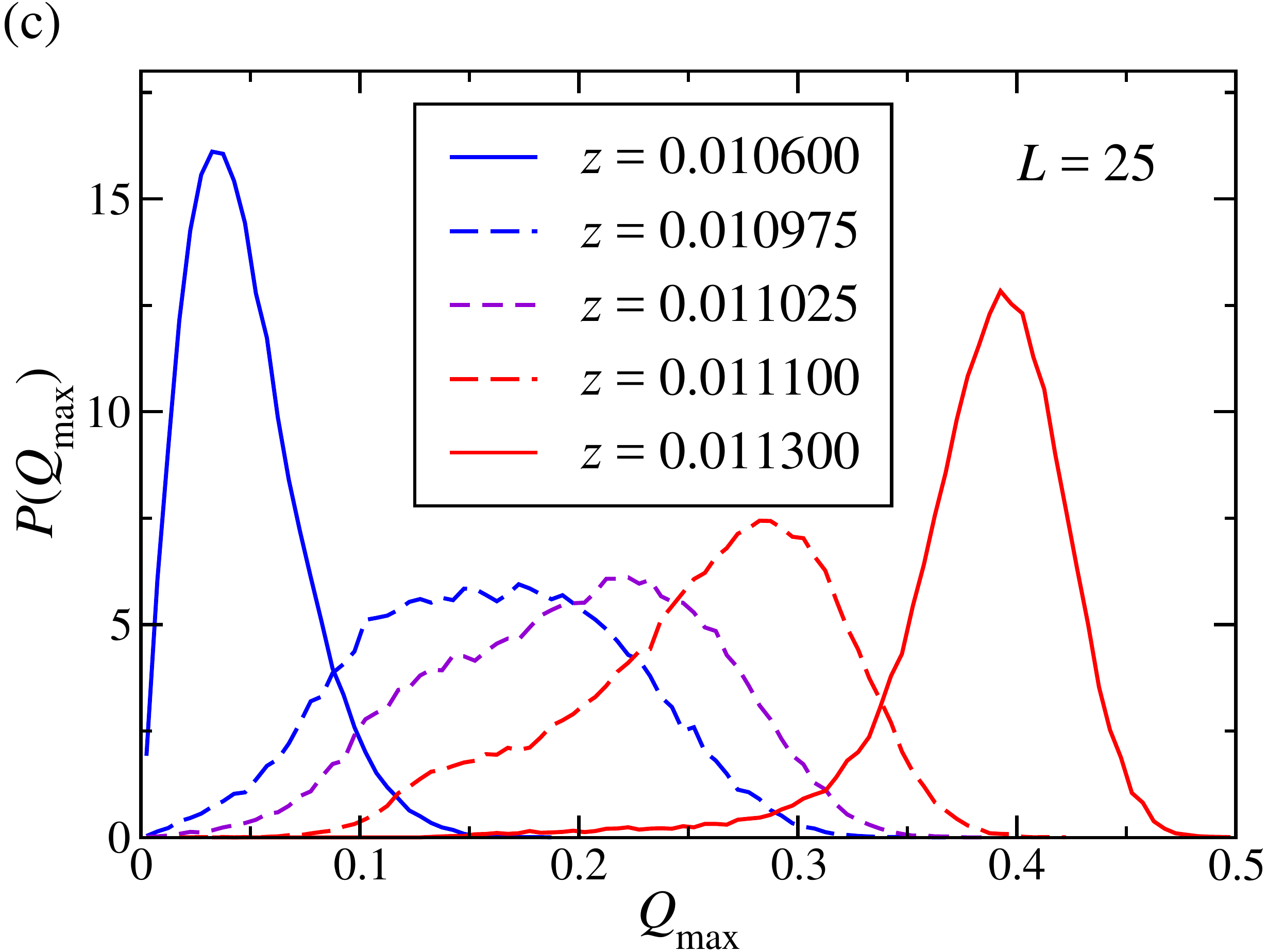}
		\label{fig:2c}
	\end{subfigure}
	\begin{subfigure}[b]{7cm}
		\phantomsubcaption
		\includegraphics[width=\textwidth]{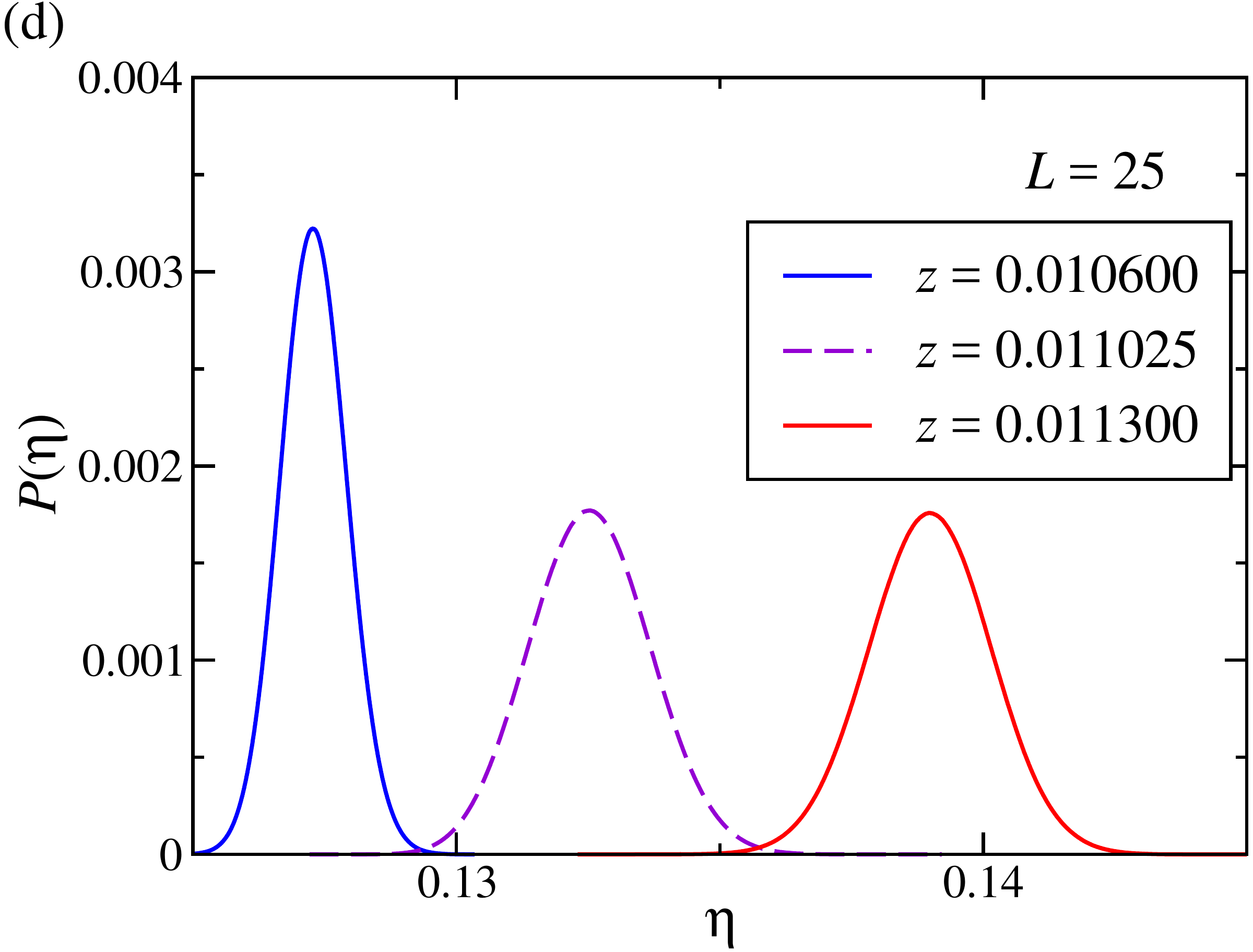}
		\label{fig:2d}
	\end{subfigure}
 \caption{(a) The distribution $P(\Qmax)$ from GCMC for $L=8$ ($M=70$, bin size $\delta Q=0.01$, measured every $5\cdot 10^7$ MC moves) 
and for several activity values between $z=0.3$ (isotropic state),
and $z=0.43333$ (nematic state). The maximal relative errors in the middle of the histograms  near coexistence ( $z=0.37000$, $z=0.38667$) are estimated at $\sim 8\%$.
(b) The corresponding distribution $P(\eta)$ for three activity values $z=0.3$ (isotropic state), $z=0.38667$ 
(near coexistence) and $z=0.43333$ (nematic state) (all measured every MC move). The maximal error estimates at the peaks of the histograms are smaller than the line thickness.
(c) The distribution $P(\Qmax)$ from GCMC for $L=25$ ($M=170$, bin size $\delta Q=0.005$, measured every $10^7$ MC moves)
and for several activity values between $z=0.0106$ (isotropic state),
and $z=0.0113$ (nematic state). The maximal relative errors in the middle of the histograms  near coexistence  are estimated at $\sim6-8\%$.
(d) The corresponding distribution $P(\eta)$ for three activity values $z=0.0106$ (isotropic state), $z=0.011025$
(near coexistence) and $z=0.0113$ (nematic state) (all measured every MC move). The maximal error estimates at the peaks of the histograms are smaller than the line thickness.}
 \label{fig:L8andL25}
\end{figure}

For $L \ge 7$, upon varying $z$, $P(\Qmax)$ transforms from having one peak near zero, via a broad two--peak structure, and finally
to a single peak with its maximum at \tr{$\Qmax >0$}. This indicates a transition from an isotropic state to a
nematic state with one excess species (positive order parameter). 
 $P(\Qmax)$ and $P(\eta)$ are shown in 
Fig.~\ref{fig:L8andL25} for rod lengths 8 and 25, among several values of $z$, respectively. The vicinity of coexistence is characterized again by two peaks in $P(\Qmax)$ smeared out broadly, yet the
corresponding peaks in $P(\eta)$ show no sign of splitting or of significant broadening. 
Therefore, the transition must entail a very small density gap, pointing to a  weak
first order transition, at most.

We additionally define the variances ${\rm Var}(\Qmax(z)) = \langle \Qmax^2 \rangle - \langle \Qmax \rangle^2$
(likewise for ${\rm Var}(\Qmin)$). 
The maximum of ${\rm Var}(\Qmax(z))$ or ${\rm Var}(\Qmin(z))$ signals the point of strongest orientational 
fluctuations, and is hence used
to locate the coexistence activity. 

\begin{figure}[h]
\includegraphics[width=7cm]{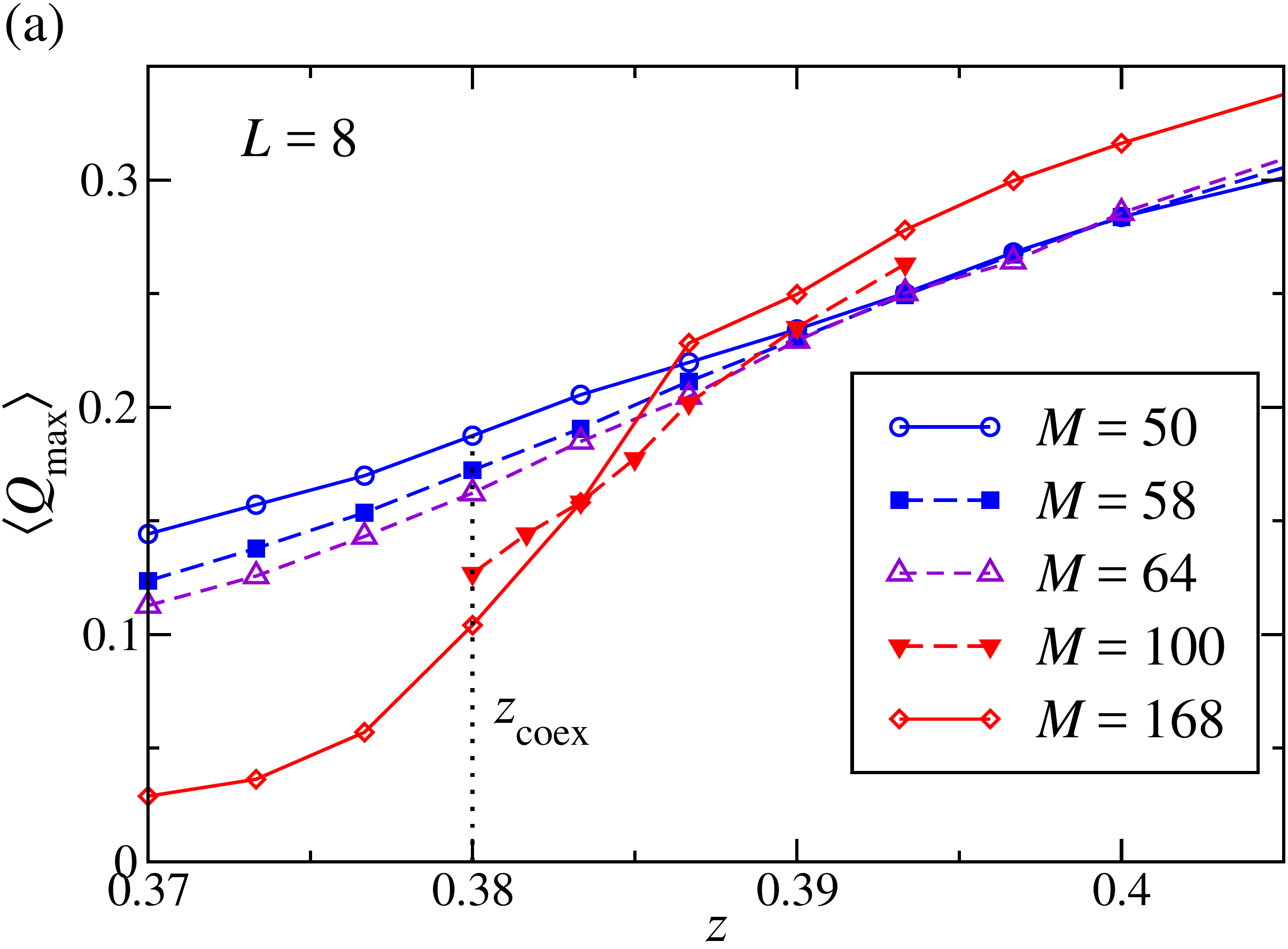} \hspace{3mm}
\includegraphics[width=7cm]{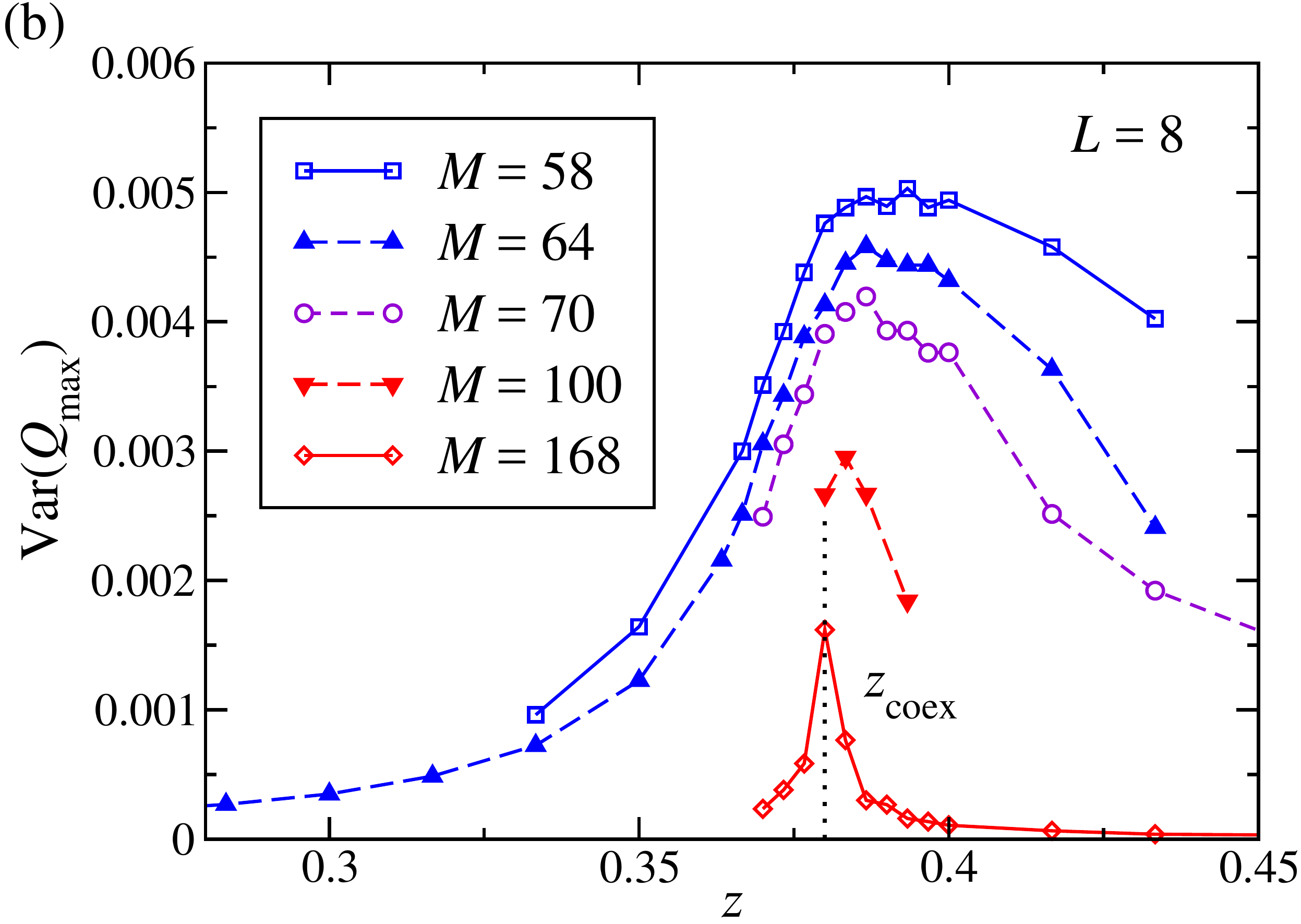}
\includegraphics[width=7cm]{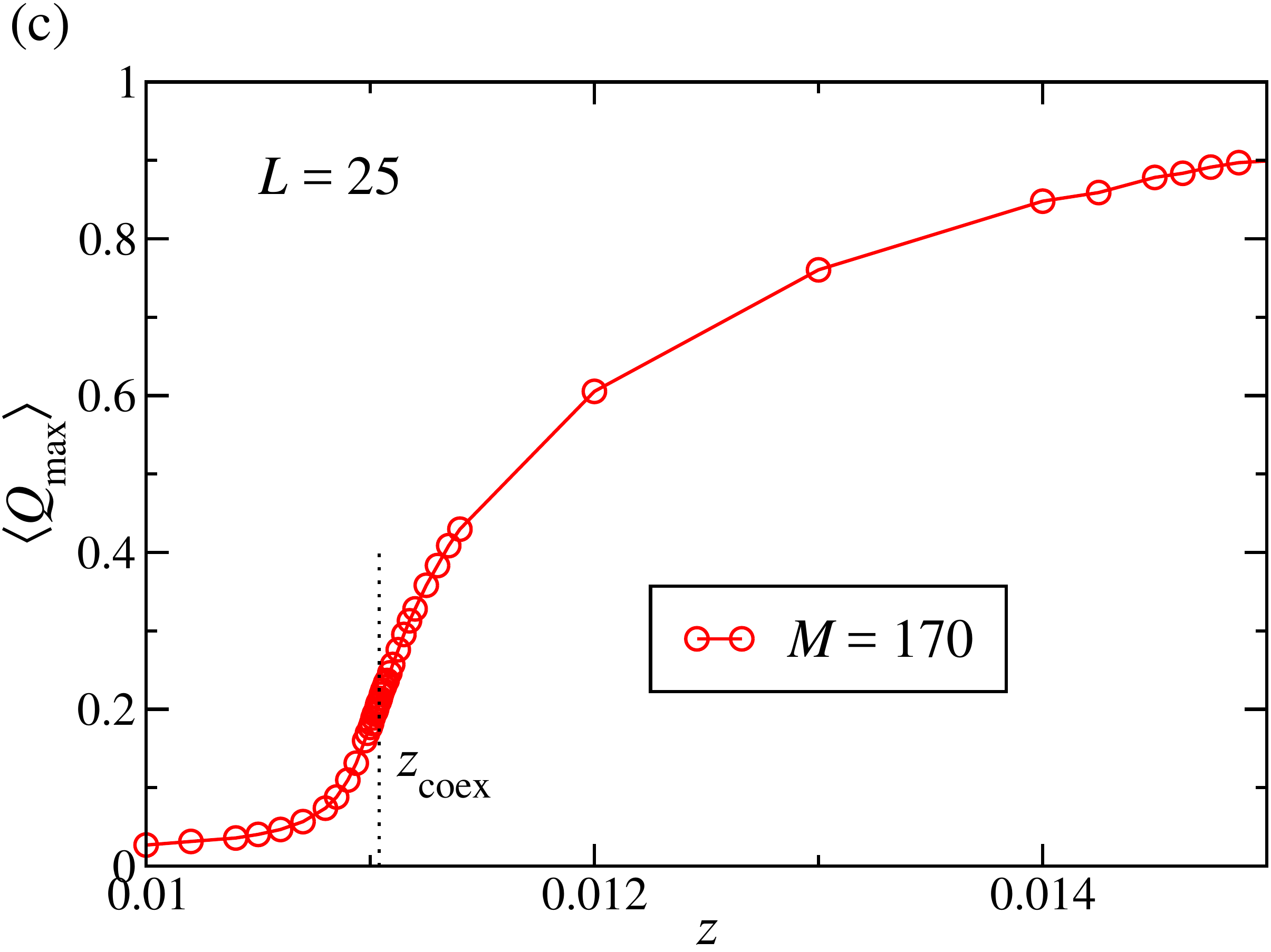} \hspace{3mm}
\includegraphics[width=7cm]{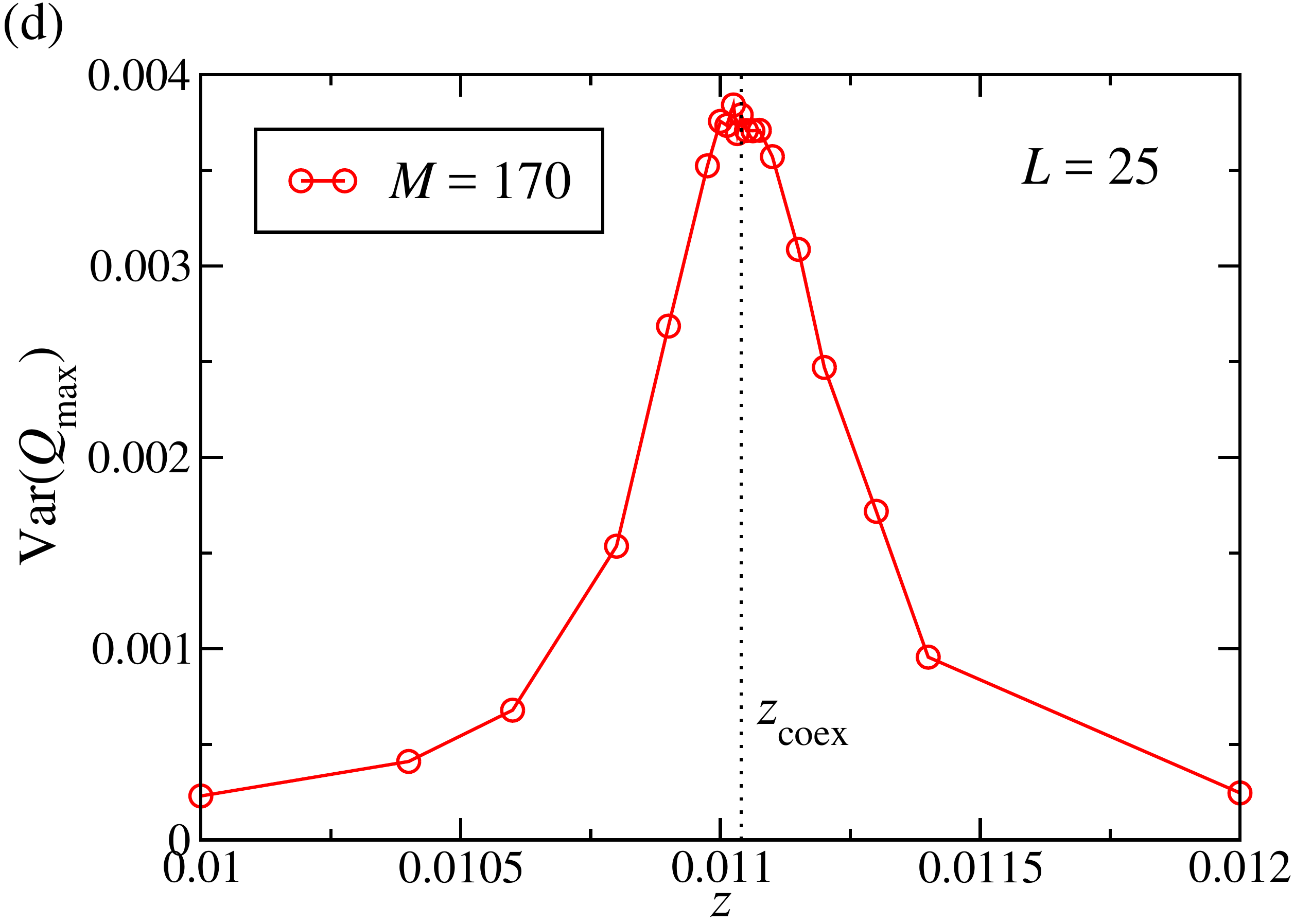}
 \caption{(a) Average value $\langle\Qmax(z)\rangle$ in the transition region from GCMC for $L=8$ and lattice extensions 
  ranging from $M=50$ to $M=168$ (measured every $5\cdot 10^7$ MC moves for $M=64,70,100$, every $10^7$ otherwise). The location of the coexistence activity $\zc$ is determined by the peak position
  of ${\rm Var}(\Qmax(z))$ for the largest lattice site. The maximal relative errors for $\langle \Qmax \rangle$ near coexistence   are estimated to less than $1\%$ for $M=168$, attributed to by the small variance (see (b)) at large system size. (b) The variance  ${\rm Var}(\Qmax(z))$ in the transition region 
  for $L=8$ and lattice extensions ranging from $M=50$ to $M=168$. (c)  Average value $\langle\Qmax(z)\rangle$ in the transition region
  for $L=25$ and lattice extension $M=170$. (d) The variance  ${\rm Var}(\Qmax(z))$ in the transition region for $L=25$.  
   }
 \label{fig:L8finitesize}
\end{figure}

For $L=8$ as an example, we have investigated finite--size effects on the behavior of $\langle\Qmax(z)\rangle$ and ${\rm Var}(\Qmax(z))$
more closely to examine their influence on the isotropic--nematic transition point. 
Fig.~\ref{fig:L8finitesize} shows these functions when varying the lattice extension from $M=50$ to 168. 
For a first order transition, $\langle\Qmax(z)\rangle$
should exhibit a jump at $\zc$, whereas  $\zc$ would be a bifurcation (critical) point for a second order transition. In any case,
a finite system size smears the jump or bifurcation. With increasing $M$, $\langle\Qmax(z)\rangle$ becomes steeper
in the coexistence region (see Fig.~\ref{fig:L8finitesize}(a)). The curves intersect at a common point for lattice extensions between 50 and 64, indicative of a second order 
transition. However,  the intersection shifts to a lower value of $z$ for $M=100$ and moreso for $M=168$,  
pointing toward a weak first order transition, consistent with the smeared two--peak structure in
$P(\Qmax)$. The corresponding behavior of ${\rm Var}(\Qmax(z))$ is shown in Fig.~\ref{fig:L8finitesize}(b):
its peak sharpens  and slightly shifts to lower $z$ for increasing $M$. The peak position for the largest
lattice extension $M=168$ is what we use to define $\zc$. The peak height in ${\rm Var}(\Qmax(z))$ shrinks with
increasing $M$. This is not quite consistent with a first--order transition for which 
the peak height should stay constant.
One sees that the average $\langle\Qmax\rangle$ at coexistence (Fig.~\ref{fig:L8finitesize}(a)) decreases for
increasing lattice size. The order parameter  is below 0.1 at coexistence for $M=168$. 
We additionally performed a finite--size analysis using two--dimensional order parameters known from three--state Potts models
\tr{and a scaling analysis for the volume susceptibility}
in Appendix \ref{sec:appendix} for the system with $L=8$. It points towards a weak first--order transition, but we cannot rule out a continuous transition with certainty.

Figures~\ref{fig:L8finitesize}(c) and (d) show $\langle\Qmax(z)\rangle$ and  ${\rm Var}(\Qmax(z))$ in the transition region
for rod length $L=25$ and large lattice extension ($M=170$). Here, the peak in the variance  ${\rm Var}(\Qmax(z))$
is readily pronounced for this lattice size, and the average value $\langle\Qmax(z)\rangle$  is around 0.2 at coexistence, i.e.
it is larger than for $L=8$, but the increase is very moderate. There is no evidence that a sizeable jump in nematic order
takes place at coexistence. 

{\em Other findings: }
Our simulations showed a particular finite-size effect that appeared for small systems (e.g. $M=32$ and $L=8$): 
the system was composed of stacked layers populated with preferably one  species, whereby the type of species varied randomly
across the layers.
We estimated that the entropic gain per particle for this layering configuration decreases inversely proportional to \tr{the area $M^2$}, 
thus this effect should vanish with larger box sizes (as confirmed by the simulations). 
We have also implemented biased sampling methods in addition to the standard GCMC algorithm. 
Specifically, we used successive umbrella sampling \cite{Vin05}, which samples equilibrium configurations strictly within a particular interval 
of a given obervable (e.g. $N$ or $Q$) via biasing---shifting successively over a range of such intervals---to obtain histograms $P(N)$ 
or $P(Q)$ with higher resolution at the tails of the distribution (the method captures the statistics of rare configurations better than standard GCMC). 
As $P(N)$ or $P(Q)$ did not show a clear signature for a first--order phase transition, however, these investigations 
did not lead to better results. 

\subsection{Comparison to FMT results}

\begin{figure}[h]
\vspace*{1cm}
\includegraphics[width=7cm]{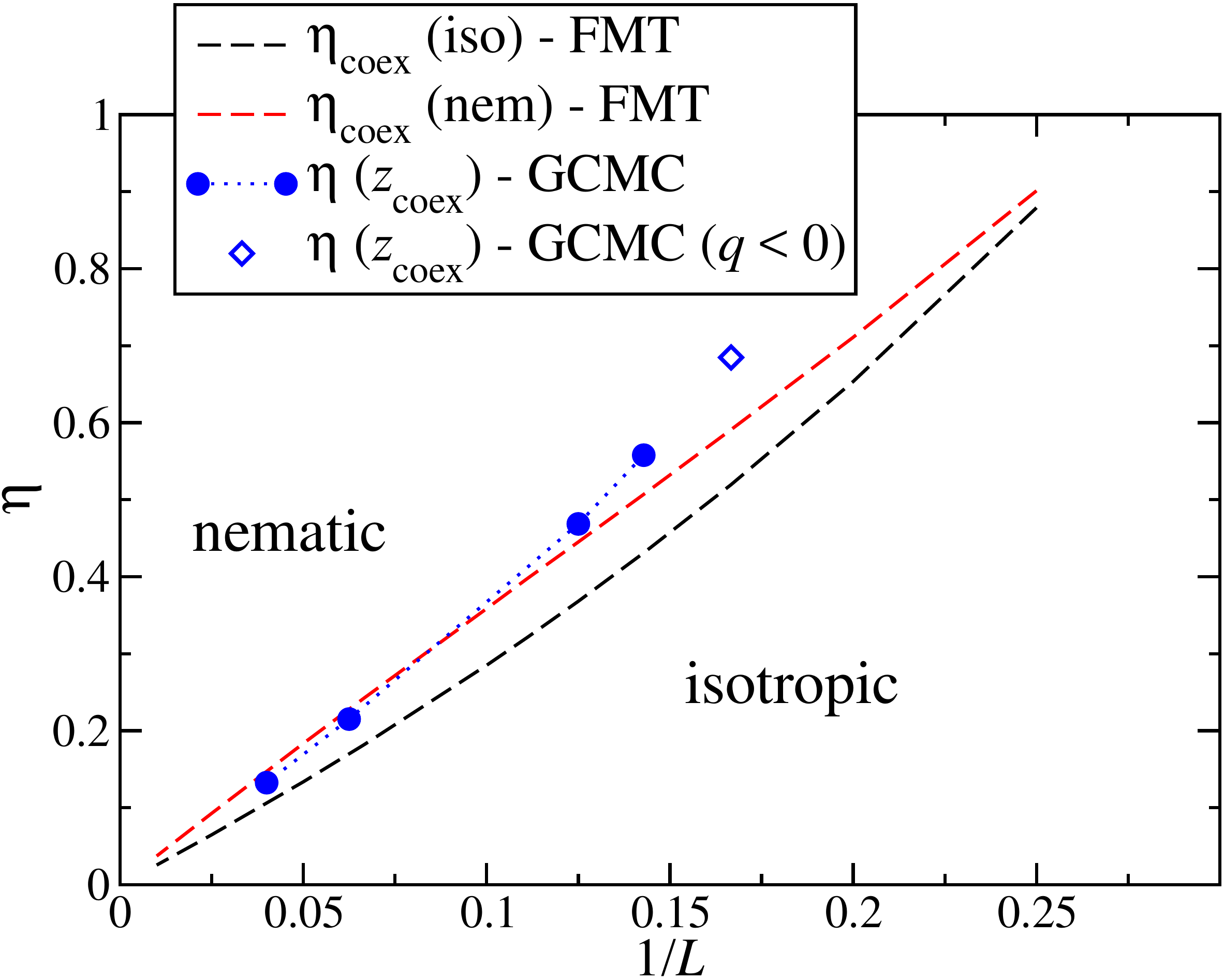} 
\caption{Phase diagram of the model in the $1/L$--$\eta$ plane. Dashed lines are FMT results (coexisting nematic states in red and coexisting
isotropic states in black). Blue symbols are GCMC results. For $L=7$, 8, 16 and 25 the nematic state has a positive order parameter 
(full symbols), whereas for $L=6$ the nematic state has a negative order parameter (open symbol).}
\label{fig:phasediagram}
\end{figure}

Lattice FMT predicts a strong first order transition for $L \ge 4$ to a nematic state with one excess species (positive order
parameter) \cite{Oet16}. We did not find stable states with a negative order parameter. The phase diagram resulting from FMT and GCMC
is shown in Fig.~\ref{fig:phasediagram}. As described before, we were not able to detect a density gap between isotropic and 
nematic states in our GCMC simulations. Surprisingly, GCMC simulations and the FMT show similar packing fractions for the 
coexistence state. We have investigated this more closely by examining 
$\langle\eta(z)\rangle$ and $\langle\Qmax\rangle(\langle\eta\rangle)$.

\begin{figure}[h]
 \includegraphics[width=7cm]{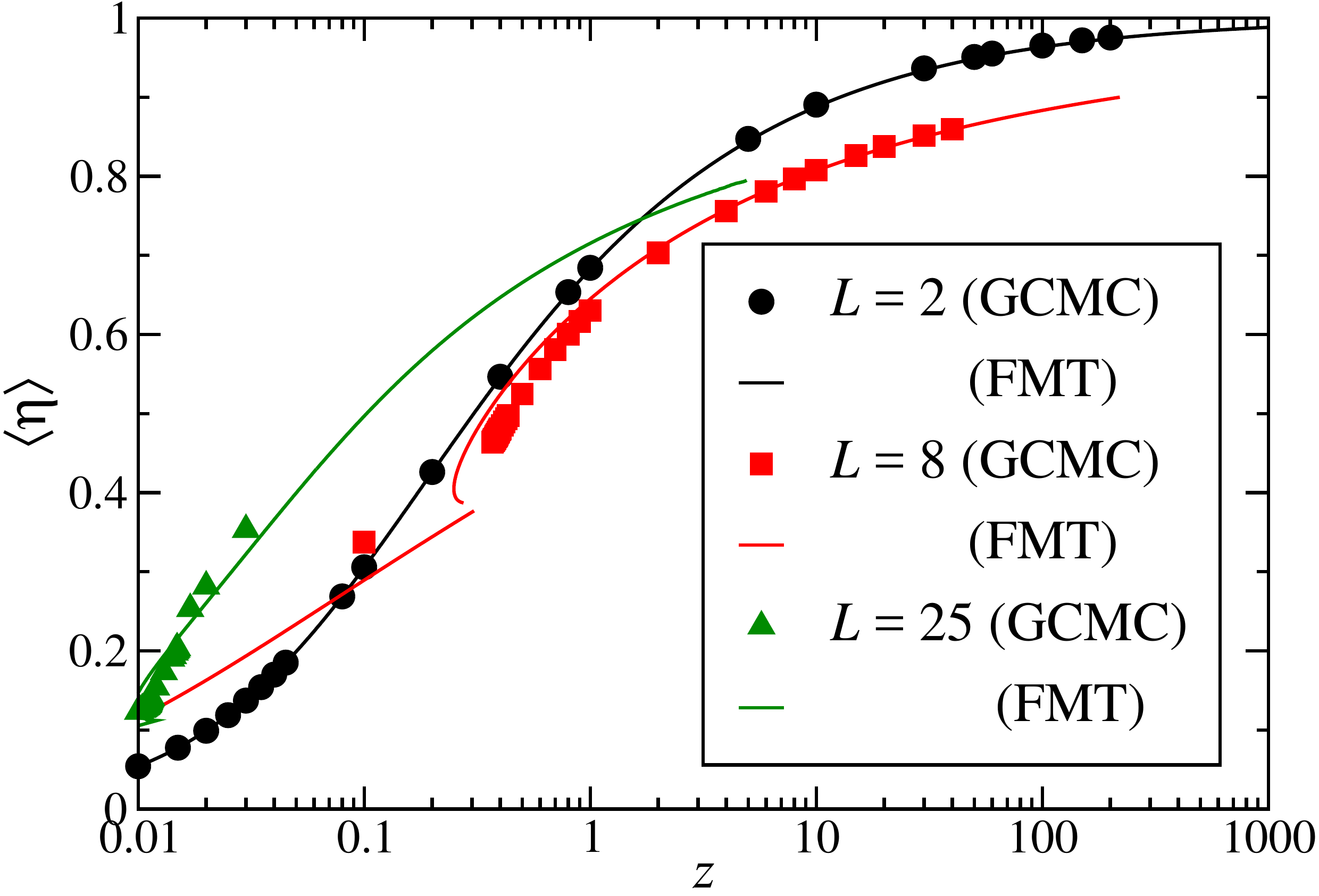}
 \caption{The expectation value of the packing fraction $\langle\eta\rangle$ as a function of the activity $z$ for rod lengths 
$L=2$, 8 and 25. Full lines are FMT results, symbols are results from GCMC simulations (error bars are smaller than the symbols). The van der Waals loop on the green curve is hidden near $z\approx 0.01$.}
 \label{fig:etaofz}
\end{figure}

The behavior of $\langle\eta\rangle(z)$ is captured very well by FMT, except for some mild disagreement in the coexistence region, see
Fig.~\ref{fig:etaofz}. A system with very short rods ($L=2$) resembles rather a hard lattice gas (no phase transition), and so the FMT results
lie on top of the GCMC data. With longer rods $L=8$ and 25 the high-- and low--density limits render good agreement with GCMC, but near coexistence the FMT data
show a van der Waals loop characteristic for a first order transition (visible for $L=8$). This is absent in the simulations.

\begin{figure}[h]
 \includegraphics[width=7cm]{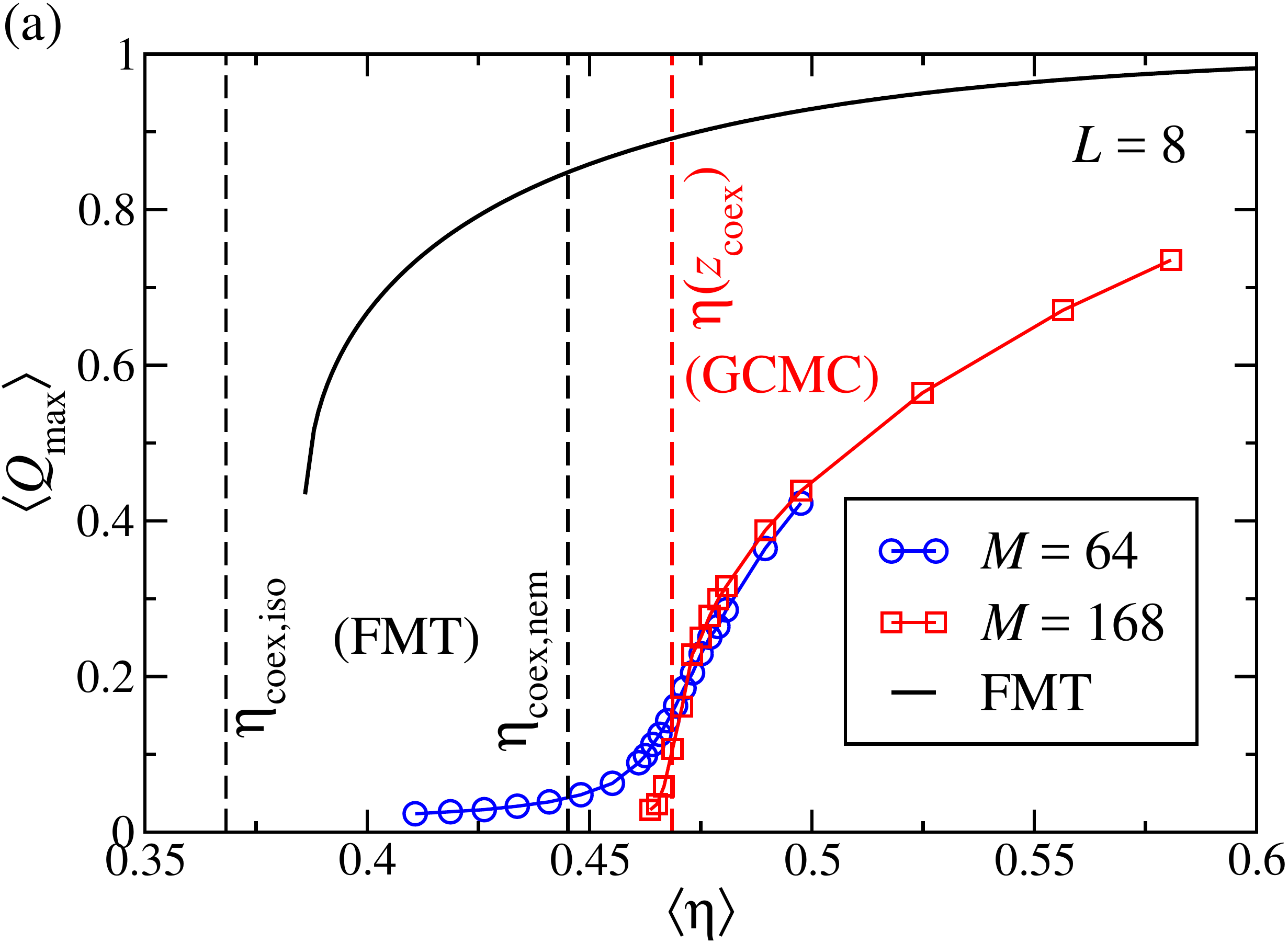} \hspace{3mm}
 \includegraphics[width=7cm]{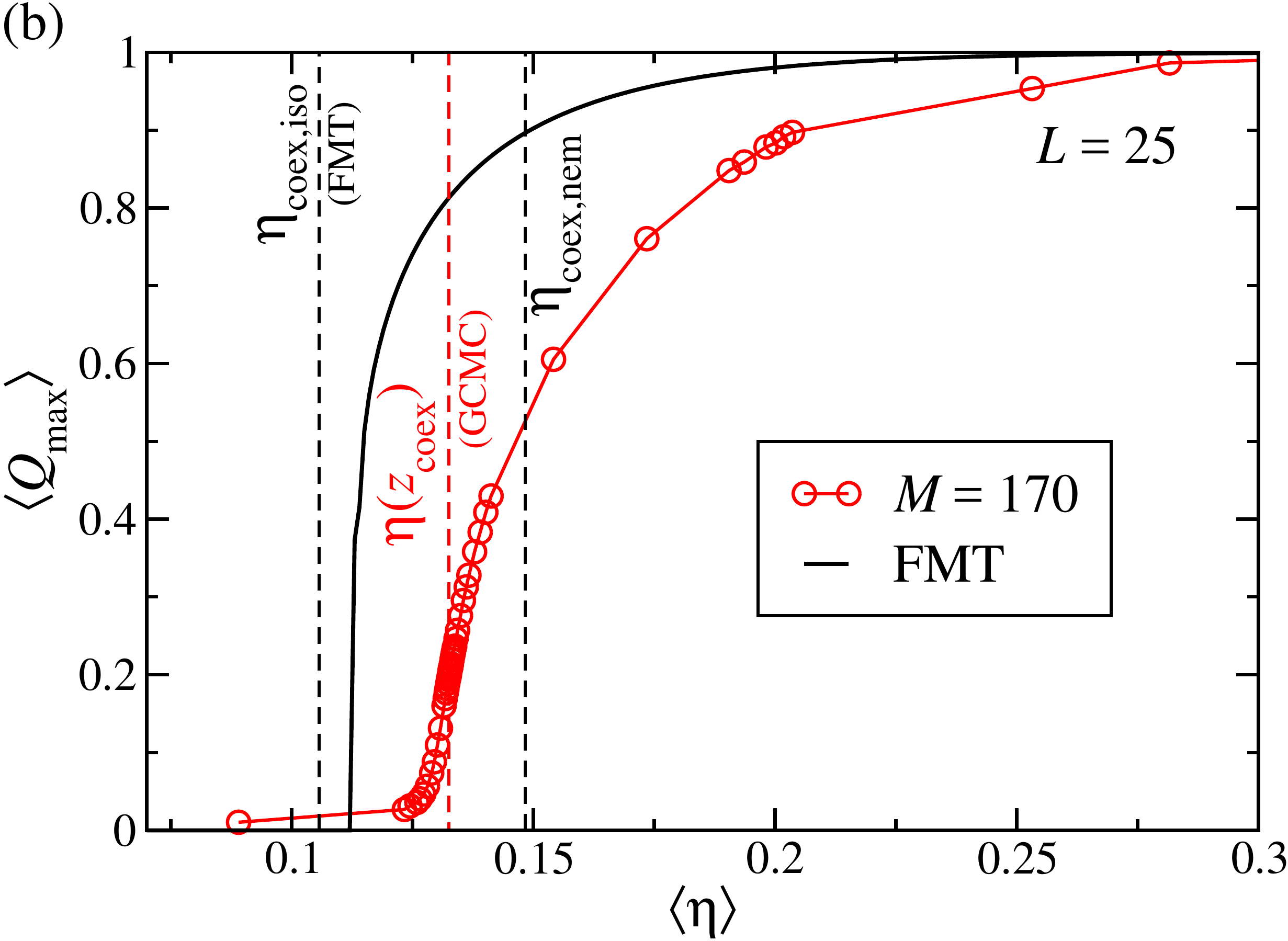}
 \caption{ The expectation value $\langle\Qmax\rangle$ as a function of the expectation value $\langle\eta\rangle$. Black lines
are FMT results, black dashed lines mark the packing fractions at coexistence from FMT and red dashed lines mark the packing fraction
at coexistence from GCMC. (a) Rod length $L=8$, symbols show GCMC results for the two lattice extensions $M=64$ (blue) and 168 (red).
(b) Rod length $L=25$, red symbols show GCMC results for a lattice extension $M=170$.
}
 \label{fig:Qofz}
\end{figure}

A side--by--side comparison of $\langle\Qmax\rangle(\langle\eta\rangle)$ unveils stronger disagreement, see Fig.~\ref{fig:Qofz}. 
For $L=8$, FMT predicts  much more pronounced 
nematic order for a given $z$. The closeness of the values for $\etac$ (GCMC) and $\etac$ of the nematic state (FMT)
thus might only be a serendipitous accident. For $L=25$, the FMT and GCMC results agree somewhat more.
It is actually not clear whether FMT and GCMC would agree in the limit $L\to\infty$ in the vicinity of the nematic transition. 
FMT renders the correct second and third virial
coefficient, and deviates from the fourth on. Although the nematic transition shifts to lower packing fraction with increasing 
$L$, higher virial coefficients affect the particular location of the transition, as  Zwanzig analyzed in the case
of a hard rod model with restricted orientational and continuous translational degrees of freedom \cite{Zwa63}. This is not the case
for hard rod models with continuous orientations, where it can be expected that a second--virial approximation is
sufficient for $L\to\infty$~\cite{Zwa63}.

\clearpage

\section{Summary and discussion}
\label{sec:summary}

Using GCMC simulations, we have found interesting characteristics of the nematic phase transition for hard rods
of size $L \times 1 \times 1$ on a 3D cubic lattice. We have observed a nematic transition for $L \ge 6$. For exactly $L=6$, the nematic state
has a negative order parameter, reflecting the occurrence of two dominating orientations. For $L \ge 7$, the nematic state
has a positive order parameter, corresponding to the dominance of one orientations. We have investigated rod lengths up to 
$L=25$ and have found evidence that the isotropic--nematic transition is of very weak first order for all rod lengths. 
It was not possible 
to  decide on the type of the transition unambiguously, 
despite using large systems containing several $10^5$ particles. The nematic order parameter
$Q$ at the transition is very small, around 0.1, and the probability distribution $P(Q)$ from the GCMC simulations are very broad, pointing to
strong fluctuations. Our results complement earlier results on the demixing (pseudonematic) transition for an equivalent system in 2D \cite{Ghosh07},
which is presumably of Ising--type and occurs for $L \ge 7$. We have compared our results to lattice FMT (which in the bulk is 
equivalent to the DiMarzio entropy \cite{DiMar61}) and find that FMT  overestimates nematic order strongly. This, however, points to
a route for how to improve the FMT. In its current form, the FMT functional is based on the exact description of 0D cavities that can hold
up to one particle. It appears sensible to extend this to cavities that can hold two or more particles. Such cavities would capture  the correlations between particles lying perpendicular to each other more
accurately, and their inclusion would also improve upon 
the mean--field assumption of DiMarzio.    

It is interesting to evaluate the use of lattice models of rods for a possible simplified description
of continuum models. We conclude our study by summarizing some findings for continuum models from the literature  and
relating these to lattice models.

{\em Continuum hard rods in 2D: }
In 2D continuum models with anisotropic particles, simulations have addressed the case of hard needles \cite{Fre85} and 
hard ellipses \cite{Fre90,Xu13}.
Since the particle orientation is indeed a continuous variable---in contrast to lattice models---fluctuations in the
average orientation may destroy the long--range order of a nematic state (leading to a quasinematic state); however,
anisotropic hard particles \tr{(for which position and orientation degrees of freedom are coupled) may still exhibit true, nematic long--range order,
see the discussion in Ref.~\cite{Fre85}.} 
Thus a continuous isotropic--quasinematic transition of Kosterlitz--Thouless type is possible, as well as an isotropic--nematic
transition of first order. Interestingly,  both types had been found in the hard ellipse system~\cite{Fre90}
(first order for aspect ratio 4 and continuous for aspect ratio 6), though, Ref.~\cite{Luo14} contradicts this and finds only continuous transitions. 
This topic may still be an open one.
Nevertheless, since the 2D lattice rod model shows an Ising--type, demixing transition,
we may conclude that in 2D lattice and continuum models
show qualitatively very different behavior with respect to the type of phase transition.

{\em Restricted-oriented rods and boards in 3D (Zwanzig model): }
Zwanzig~\cite{Zwa63} initiated investigating the packing of hard rods with 3D mutually--orthogonal orientations (a restricted--orientations model) 
in his seminal work. For long rods, he found that the virial coefficients scaled very differently  with aspect ratio compared to those of hard rod models with
unrestricted orientations. Therefore, a second--virial approximation is not sufficient to locate the isotropic--nematic transition.
This finding should similarly apply to the 3D lattice model investigated by us.   
Martinez--Raton~\cite{Mart04} calculated the full phase diagram using an approximate FMT. \tr{There, a nematic state with negative
order parameter is found in conjunction with layering. Thus the appearance of such nematic states in \cite{Mart04} and in the present work
seems to be a direct consequence of restricting particle orientations. }

{\em Boards and cuboids in the 3D continuum: }
These models release the restriction on orientation, but retain the particle anisometry and shape of the aforementioned Zwanzig models.
Refs.~\cite{Esco05,Esco08} simulated bulk properties of hard tetragonal parallelpipeds, 
havign found a parquet phase for the case of cuboids with two short, symmetrical axes (i.e. rods). 
This is absent in the Zwanzig models. Ref.~\cite{Perou13} investigated more deeply the
appearance of biaxial phases in these systems.

{\em Continuum hard rods (spherocylinders) in 3D: }
Spherocylinders have proven to be the most sought-after model with regards to hard anisotropic particles in the 3D continuum.
On the simulation side, Bolhuis and Frenkel~\cite{Bol96diss,Bol97} used MC simulations with $\sim500$ particles in the canonical ensemble.
The nematic phase was found to be stable for aspect ratios $L/D \ge 3.8$.
The isotropic--nematic transition has been assumed to be first order but for small aspect ratios
$L/D \le 5.0$ it was not possible to detect a density jump between coexisting states. 
The order parameter $Q^{\rm nem}(\rho)$ showed weak hysteresis, but a jump in $Q^{\rm nem}$ between
coexisting states (a clear signature of a first order transition) was not determined.
The transition density is estimated from
the kink in the $Q^{\rm nem}$--$\rho$--curve at the (arbitrary) value $Q^{\rm nem}=0.4$. For $L/D \ge 15$
a density jump of $>10\%$ had been found, the absolute value of the density jump was maximal at around $L/D=20$.
The nematic order parameter of the nematic phase at coexistence was found to be 0.784 at $L/D=15$ and further increases
when the aspect ratio lengthens. So, for $L/D \ge 15$, the transition is very strongly first order.
Vink et al. \cite{Vin05} employed grand canonical MC techniques. For $L/D=15$ they confirmed the sizable density jump, and at
coexistence the probability distribution $P(\rho)$ showed two clearly separated peaks.  

On the theory side,
the most accurate density functionals for hard anisotropic particles have been derived from Fundamental Measure Theory (FMT)
\cite{Han09,Han10,Witt14a,Witt14b,Witt15,Witt16}. Wittmann et al. used mixed measures to derive a functional that is exact in the
low--density limit; for higher densities, typical FMT approximations were employed \cite{Witt15}.
The corresponding phase diagram for hard spherocylinders showed almost quantitative agreement with Ref.~\cite{Bol97}.
However, the FMT results showed an almost constant density jump between the coexisting isotropic and nematic states
for aspect ratios starting from $L/D \gtrsim 3.5$. Thus the transition is strongly first order for all aspect ratios.  

It hence appears that for moderate aspect ratios strong orientational fluctuations cause the isotropic--nematic transition
to be weakly first order, both on the lattice and in the continuum. The lattice model exhibits in addition the peculiarity of a nematic
state with negative order parameter (for $L=6$). The lattice and the continuum stand in contrast to each other for large aspect ratios  since we have not found
evidence for a strong first order transition. Moreover, it does not seem that a smooth crossover in the
topology of the phase diagram is possible from the lattice to the continuum: this was shown by a study using a second--virial density functional with variable orientational degrees of freedom~\cite{Shu04}.    

In the present work, we have not investigated the possibility of \tr{high--density phases} in this system,
\tr{owing to limitations of our simple grand--canonical algorithm. These high--density phases could 
include a completely disordered phase} of cubatic--type, which could be similar
to the high--density phase in 2D \cite{Ghosh07,Kun13}. This problem should be treated with optimized algorithms as has been done in
2D \cite{Kun13}. 

\tr{{\bf Note added:} After submission of this work, a study on the same system was published
on arXiv \cite{Raj17}. There, the authors used optimized algorithms for higher densities and could show
that the layered nematic phase with $q<0$ not only appears for $L=6$ but also for $L=5$ and $L=7$ at high packing fractions
of around 0.9. The corresponding findings for the isotropic--nematic($q>0$) transition are similar to ours. The systems studied
were smaller than ours, therefore this transition appeared to be critical in the size regime used.      } 

{\bf Acknowledgments: } 
We thank Richard Vink (G\"ottingen) for very valuable discussions.
This work is supported within the DFG/FNR INTER project 
``Thin Film Growth" by the Deutsche Forschungsgemeinschaft (DFG), project OE 285/3-1, and by the Landesgraduiertenf\"orderung Baden--W\"urttemberg.
Y. Ai thanks the German Academic Exchange Service (DAAD) for a RISE grant in summer 2015.

\clearpage

\begin{appendix}
\section{Finite size analysis for $L=8$}
\label{sec:appendix}

\begin{figure}[h]
 \includegraphics[width=5cm]{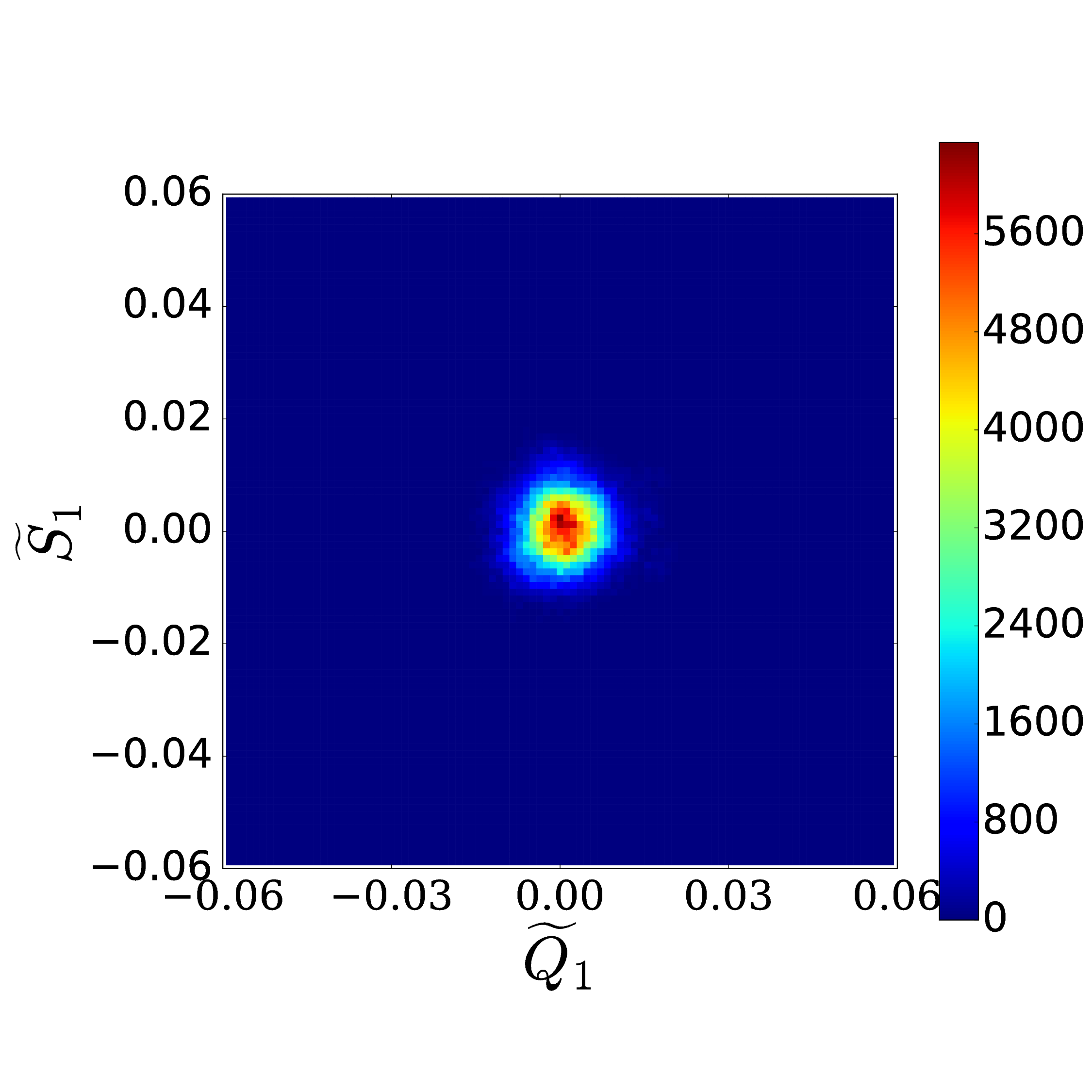}
 \includegraphics[width=5cm]{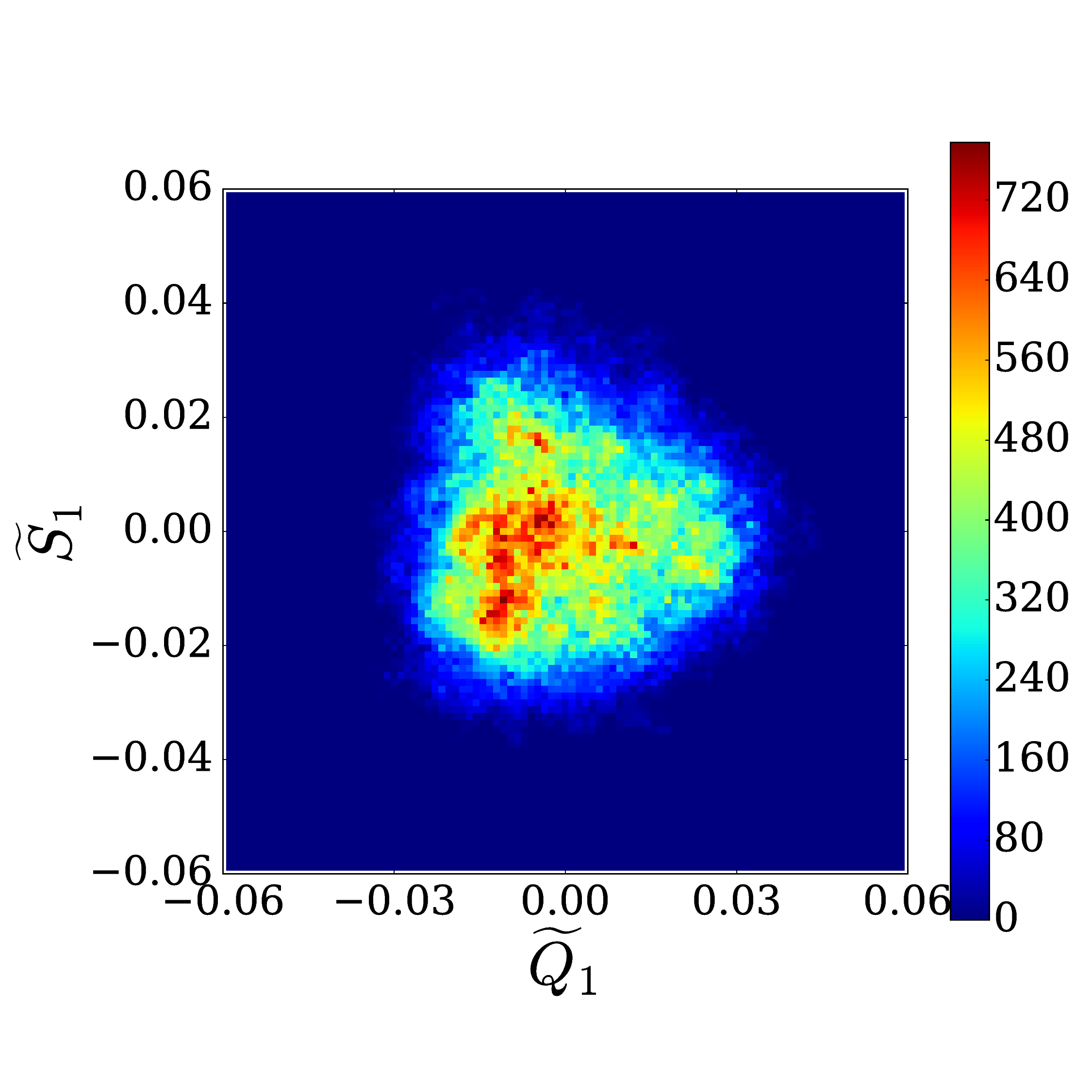}
 \includegraphics[width=5cm]{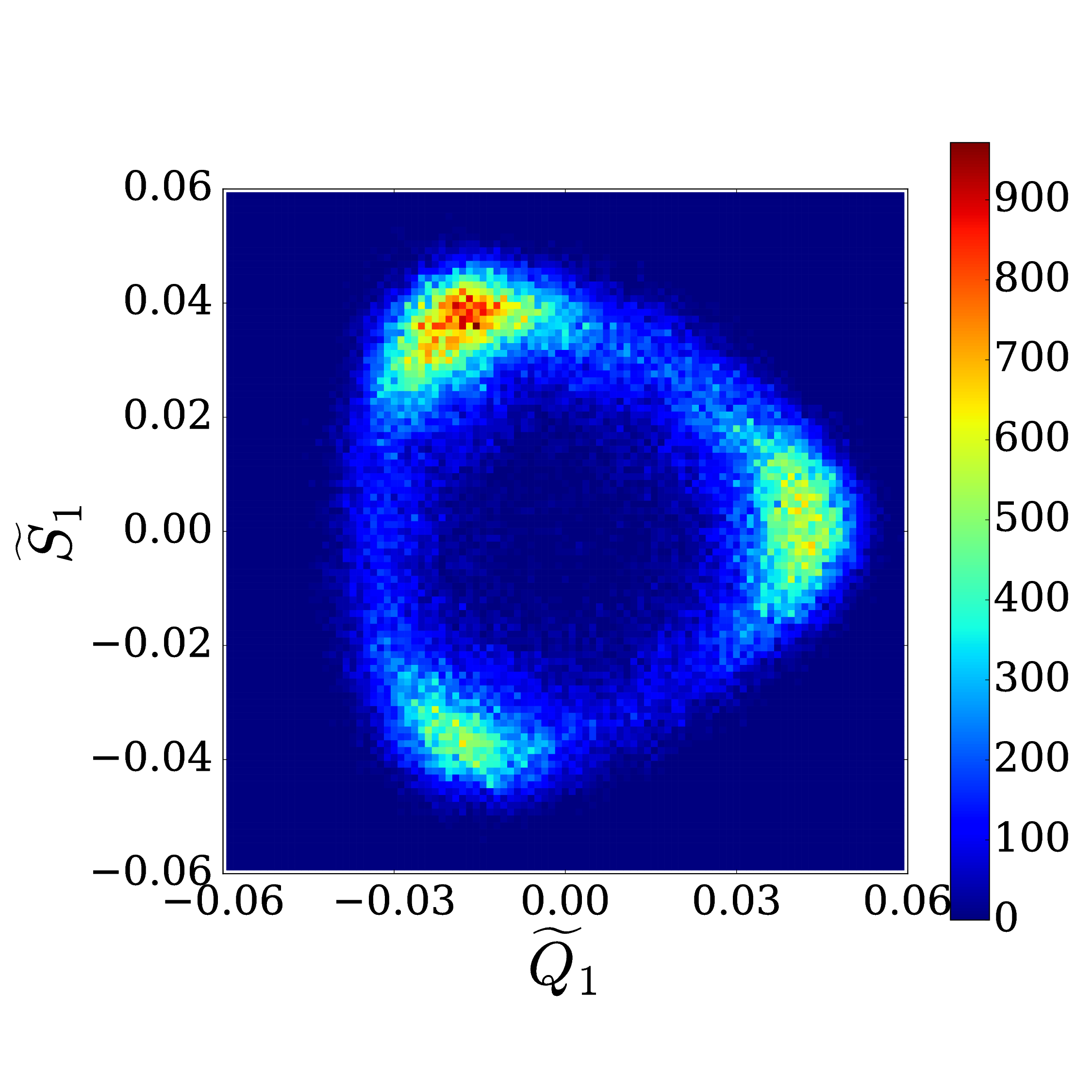}
 \caption{ 2D histograms $P(\tilde{S_1}, \tilde{Q_1})$ for three states: isotropic ($z=0.35$), near the transition ($z=0.37667$), and above
 the transition in the nematic state ($z=0.38667$), as obtained for $L=8$ and $M=100$.
}
 \label{fig:2Dhisto}
\end{figure}

Here we apply a general approach for constructing multidimensional order parameters, developed for $q$--state Potts models \cite{Bind93}.
The discussion in \cite{Bind93}, done in the canonical ensemble, can be mapped directly to the purely entropic
lattice model treated in the grand--canonical ensemble. This can be done by identifying the temperature with the chemical potential, and the
spin states with the rod species.
Regarding the symmetries of the order parameter, our lattice model is equivalent to a three--state Potts model 
where the symmetry--broken phase 
consists of three equivalent states (each of the Cartesian axes can be a preferred direction).
According to Ref.~\cite{Bind93}, the order parameter dimensionality is two, and 
orthogonal axes in order parameter space are formed by pairs $(\tilde Q_i, \tilde S_i)$ of {\em unnormalized} nematic and biaxial
order parameters (Eqs.~(\ref{eq:Qidef}) and (\ref{eq:Sidef})):
\bea
 \label{eq:tildeQidef}
   \tr{\tilde Q_i} & = &  {\rho_i - \frac{\rho_j +\rho_k}{2}} \;, \\
 \label{eq:tildeSidef}
   \tr{\tilde S_i} &=  &  \frac{\sqrt{3}}{2}(\rho_j -\rho_k) \;,
\eea 
where $(ijk)$ is a cyclic permutation of $(123)$. In 2D histograms $P(\tilde{S_i}, \tilde{Q_i})$, a single peak around the origin signals 
an isotropic state, whereas three peaks with the same distance from the origin and with a threefold symmetry signal a nematic phase. 
The simultaneous occurrence of all four peaks would correspond to nematic--isotropic coexistence. In Fig.~\ref{fig:2Dhisto},
we show $P(\tilde{S_i}, \tilde{Q_i})$ for three values of $z$ in a system with rod length $L=8$ and lattice extension
$M=100$. While the isotropic and nematic state can be clearly identified, the histogram near the transition (middle graph in
Fig.~\ref{fig:2Dhisto}) shows a broad distribution. The four peaks (expected for coexistence) cannot be identified clearly.

\begin{figure}[h]
 \includegraphics[width=7cm]{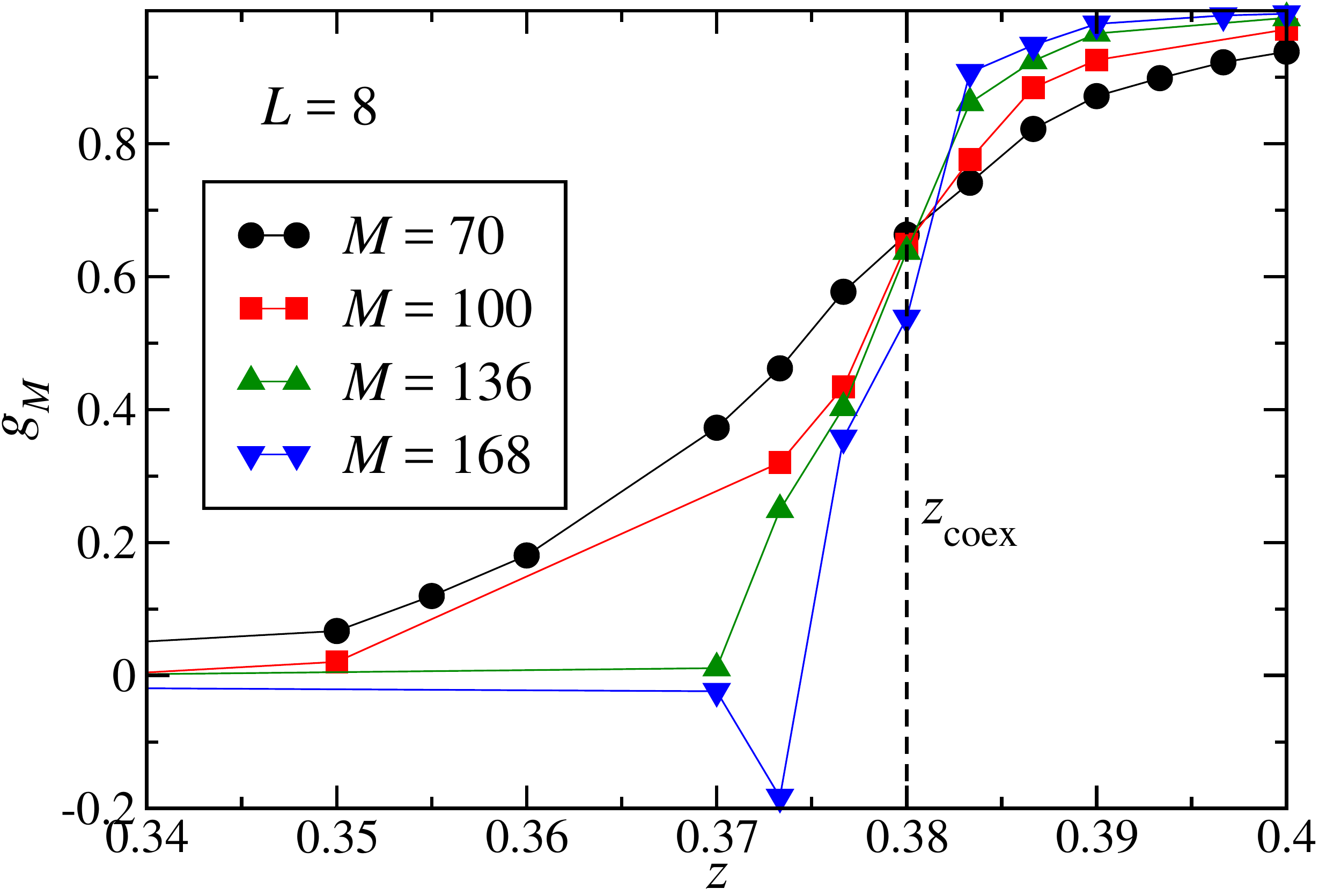}
 \caption{ The reduced (Binder) cumulant $g_M$ in a system with $L=8$ for different lattice extensions $M$.
}
 \label{fig:BinderCumulant}
\end{figure}

In order to facilitate a distinction between first order and continuous transitions, Ref.~\cite{Bind93} introduces the reduced
cumulant
\bea
 g_M = 2 - \frac{\langle m_i^4 \rangle}{\langle m_i^2 \rangle^2} \;,
\eea 
where $m_i$ is the radial distance in the order parameter plane, $m_i = \sqrt{ \tilde Q_i^2 +\tilde S_i^2}$. 
Cumulants for different lattice extensions $M$ will intersect at $\zc$. In the case of a first order transition,
$g_M$ will develop a minimum for increasing $M$ below $\zc$, and $\zc-z_{\rm min}$ (where $z_{\rm min}$ is the position
of the minimum) scales as $1/M^3$. In Fig.~\ref{fig:BinderCumulant} we show $g_M$ for the system with $L=8$. The 
common intersection at $z \approx 0.38$ confirms the transition activity identified via the position of the maximum
in Var($\Qmax$) for the largest system ($M=168$). However, our largest system $M=168$ is just large enough to see
an indication of a minimum in $g_M$; therefore, we cannot do a finite size analysis for the position of the minimum. Nevertheless, the
existence of the minimum would point toward a weak first order transition. Investigations much more extensive  than the present
one would be needed to establish the order of the transition unambiguously.  

\tr{
A scaling analysis corroborates the transition to noncritical behavior for large system sizes. 
Consider the unnormalized volume susceptibility $\xi_{\rm max} = {\rm Var}(\max\limits_{(ijk)}\{N_i-(N_j+N_k)/2\})$ which
corresponds to the normalized version expressed by ${\rm Var}(\Qmax)$. For a second order transition, one would expect
to find a scaling function $\tilde\xi(x)$ with the scaling variable $x=(1- z/\zc)M^{1/\nu}$ and 
the relation $\tilde\xi(x)=M^{-3-\gamma/\nu} \xi(x)$ \cite{Barkema}. Here $\gamma$ and $\nu$ are critical exponents
in standard notation. The such rescaled susceptibilities for $L=8$ and for different $M$ are shown in 
Fig.~\ref{fig:scaling} using 3D Ising critical exponents ($\gamma \approx 1.237, \nu \approx 0.630$).
The critical exponents of the 3D antiferromagnetic Potts model on a cubic lattice are similar \cite{Kol95}. 
For lattice extensions $M=64...136$, the data collapse on a single scaling curve for $x<0$ (isotropic side).
For $x>0$ (nematic side) the collapse is not perfect with deviations coming mainly from small systems.
However, the data for $M=168$ clearly deviate from the approximate scaling function, indicating that
the transition is not critical anymore. This matches very well with the conclusions drawn from the
cumulant analysis.  
\begin{figure}[h]
 \includegraphics[width=7cm]{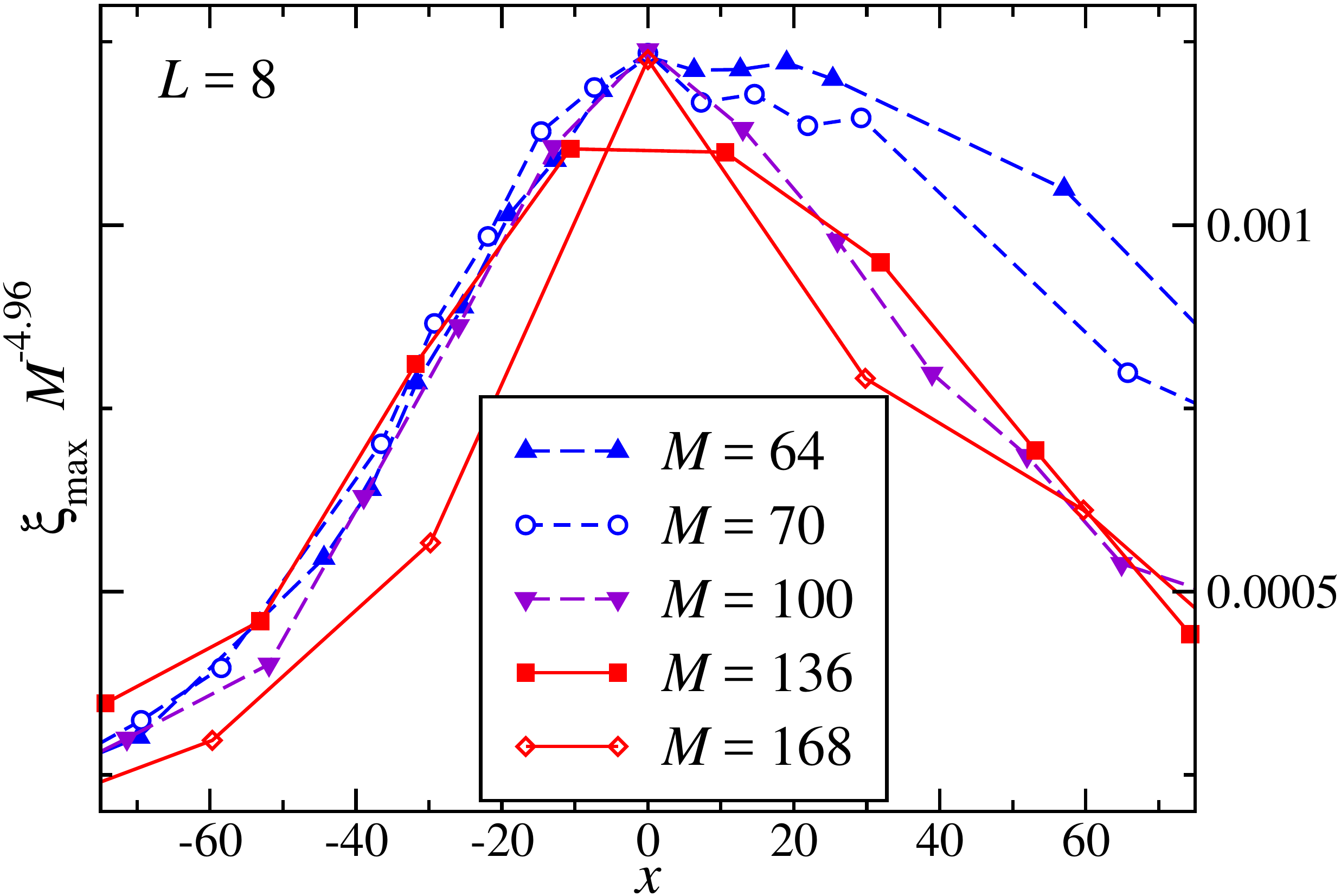}
 \caption{ \tr{Rescaled volume susceptibilities $M^{-3-\gamma/\nu} \xi(x)$ ($x=(1- z/\zc)M^{1/\nu}$) 
    for $L=8$ and for different lattice extensions.}
}
 \label{fig:scaling}
\end{figure}
}
\end{appendix}

\end{document}